\documentclass[a4paper, 10pt]{article}

\usepackage[english]{babel}
\usepackage{latexsym}
\usepackage{dsfont}       %Doppelbalkenbuchstaben \mathds{}
\usepackage{graphicx}
\usepackage[latin1]{inputenc} %Einfache Eingabe von Umlauten
\usepackage[T1]{fontenc}  %Trennung von Umlauten enthaltenden Wörtern (wenn im PDF die Schrift übel aussieht:)
\usepackage{mathrsfs}
\usepackage{bm}           % für fette schrift in formeln...
\usepackage{array}
\usepackage{caption2}     % um captions bei bildern zu gestalten

\usepackage{amsmath, amssymb} % for handling labels, endmarks...
\usepackage[amsmath, thmmarks]{ntheorem} % um massgerechte theorem-umgebungen zu definieren. brauche amsmath einzufüghen, um kompatibel mit \usepackage{amsmath} zu sein (= standard LaTeX theorem umgebung). brauche thmmarks einzufügen um endmarks einzufügen. Falls kein endmark erwünscht, jeweils \NoEndMark nach \begin{Prop}... hinschreiben.

\newcommand{\pre}{\theorempreskipamount}
\newcommand{\post}{\theorempostskipamount}
\theoremstyle{break}
\theoremheaderfont{\normalfont\bfseries}
\theorembodyfont{\itshape}                                                      % itshape - italic shape, i.e. kursiv
\theoremseparator{\;\;}
\newtheorem{thm}{Theorem}[section]              % \newtheorem*{}[] gibt theorem umgebung welche nicht nummeriert ist

\theoremstyle{plain}
\theoremheaderfont{\normalfont\bfseries}
\theorembodyfont{\itshape}
\theoremseparator{\;\;}

\theoremstyle{plain}
\theoremheaderfont{\normalfont\bfseries}
\theorembodyfont{\itshape}
\theoremseparator{\;\;}
\newtheorem{lem}[thm]{Lemma}

\theoremstyle{break}
\theoremheaderfont{\normalfont\bfseries}
\theorembodyfont{\normalfont}
\theoremsymbol{\ensuremath{\Box}}
\theoremseparator{\;\;}
\newtheorem{model}[thm]{Model Assumptions}

\theoremstyle{break}
\theoremheaderfont{\normalfont\bfseries}
\theorembodyfont{\normalfont}
\theoremsymbol{\ensuremath{\Box}}
\theoremseparator{\;\;}
\newtheorem{algorithm}[thm]{Algorithm}

\theoremstyle{plain}
\theoremheaderfont{\scshape}
\theorembodyfont{\normalfont}
\theoremsymbol{\ensuremath{\Box}}
\theoremseparator{\;\;}
\newtheorem{dfn}[thm]{Definition}

\theoremstyle{plain}
\theoremheaderfont{\scshape}
\theorembodyfont{\normalfont}
\theoremsymbol{\ensuremath{\Box}}
\theoremseparator{\;\;}
\theoremnumbering{arabic}
\newtheorem{bsp}[thm]{Example}

\theoremstyle{plain}
\theoremheaderfont{\scshape}
\theorembodyfont{\normalfont}
\theoremsymbol{\ensuremath{\Box}}
\theoremseparator{\;\;}
\theoremnumbering{arabic}
\newtheorem{rmk}[thm]{Remark}

\theoremstyle{break}
\theoremheaderfont{\scshape}
\theorembodyfont{\normalfont}
\theoremsymbol{\ensuremath{\Box}}
\theoremseparator{:}
\theoremnumbering{arabic}
\newtheorem{rmks}[thm]{Remarks}

\theoremstyle{nonumberplain}
\theoremheaderfont{\normalfont\bfseries}
\theorembodyfont{\normalfont}
\theoremseparator{\;\;}
\newtheorem{update}{Information update process.}

\theoremstyle{change}
\theoremheaderfont{\sc}
\theorembodyfont{\normalfont}                   % slshape, upshape
\theoremseparator{:\;\;}
\theoremsymbol{\ensuremath{\Box}}  % variante: \rule{1.2ex}{1.2ex} würde eigenes Symbol für proof umgebung kreiren: schwarze box.
\newtheorem*{Proof}{Proof}

\theoremstyle{change}
\theoremheaderfont{\sc}
\theorembodyfont{\normalfont}
\theoremseparator{:\;\;}
\theoremsymbol{\ensuremath{\Box}}

\newcommand{\be}{\begin{equation}}
\newcommand{\ee}{\end{equation}}
\newcommand{\bd}{\begin{displaymath}}
\newcommand{\ed}{\end{displaymath}}

\numberwithin{equation}{section}
\title{The Quantification of Operational Risk using Internal Data, Relevant External Data and Expert Opinions}
\author{

Dominik D.~Lambrigger\footnotemark[1] \quad Pavel V.~Shevchenko\footnotemark[2] \quad Mario V.~Wüthrich\footnotemark[1]}

\date{First version: April 13, 2007\\
This version: July 4, 2007}
\begin{document}
\maketitle

\footnotetext[1]{ETH Zurich, Department of Mathematics, CH-8092 Zurich, Switzerland.}
\footnotetext[2]{CSIRO Mathematical and Information Sciences, Sydney, Locked Bag 17, North Ryde, NSW, 1670, Australia; e-mail: Pavel.Shevchenko@csiro.au.}

\begin{center}
This is a preprint of an article published in \\The Journal of
Operational Risk 2(3), pp.3-27, 2007.\\
www.journalofoperationalrisk.com
\end{center}

\begin{abstract}
\noindent
To quantify an operational risk capital charge under Basel II, many banks adopt a Loss Distribution Approach. Under this approach, quantification of the frequency and severity distributions of operational risk involves the bank's internal data, expert opinions and relevant external data. In this paper we suggest a new approach, based on a Bayesian inference method, that allows for a combination of these three sources of information to estimate the parameters of the risk frequency and severity distributions.
\end{abstract}


\section{Introduction}
To meet the Basel II requirements, BIS \cite{Basel}, many banks adopt a Loss Distribution Approach (LDA). Under this approach, banks quantify distributions for the frequency and severity of operational losses for each risk cell over a one year time horizon; see, e.g., Cruz \cite{Cruz}, McNeil et al. \cite{MFE}, Panjer \cite{Panjer}. Banks can use their own risk cell structure but they must be able to map the losses to the relevant Basel II risk cells (eight business lines times seven risk types). The commonly used LDA model for an annual loss in a single risk cell is the sum of individual losses
\be
L = \sum_{k=1}^{N} X_k,
\ee
%Thereby $N$ is the annual number of events (frequency) modeled as a random variable (rv) from some discrete distribution $F_N(\cdot|\bm{\gamma})$. $X_k$, $k=1,\ldots,N$, are the severities of the events modeled as independent rvs from a continuous distribution $F_X(\cdot|\bm{\gamma})$. The vector $\bm{\gamma}$ represents the underlying parameter distribution, giving each company its (unknown) specific risk characteristics $\bm{\gamma}_0$.\\
where $N$ is the annual number of events (frequency) and $X_k$, $k=1,\ldots,N$, are the severities of these events.\\
Several studies, e.g., Moscadelli \cite{Mo04} and Dutta and Perry \cite{DuPe06}, analyzed operational risk data collected over many banks by Basel II business line and event type; see Degen et al.~\cite{DEL} for a discussion and analysis of these studies. While analyses of collective data may provide a picture for the whole banking industry, estimation of frequency and severity distributions of operational risks for each risk cell is a challenging task for a single bank. The bank's internal data are typically collected over several years. On the one hand, there might be some cells with few internal data only. On the other hand, industry data available through external databases (from vendors and consortia of banks) are often difficult to adapt to internal processes, due to different volumes, thresholds etc.\\
% and may contain few losses or even none for some risk cells. The industry data are available through external %databases but these are difficult to use directly due to different volumes and other factors. Moreover, the data have %a survival bias as typically the data of all collapsed companies are not available. It is difficult to estimate %distributions using these data only. It is also clear that this estimation is backward looking and has limited ability %to predict the future due to a constantly changing banking environment.\\
Therefore, it is important to have expert judgments incorporated into the model. These judgments may provide valuable information for forecasting and decision making, especially for risk cells lacking internal loss data. In the past, quantification of operational risk was based on such expert judgments only. A quantitative assessment of risk frequency and severity distributions can be obtained from expert opinions; see, e.g., Alderweireld et al.~\cite{AlGaLe06}. By itself, this assessment is very subjective and should be combined with (supported by) the analysis of actual loss data.
In practice, due to the absence of a sound mathematical framework, ad-hoc procedures are often used to combine the three sources of data: internal observations, external data and expert opinions. For example, the frequency distribution is estimated using internal data only, while the severity distribution is fitted to a sample combining internal and external data.\\
On several occasions, risk executives have emphasized that one of the main challenges in operational risk management is to combine internal data and expert opinion with relevant external data in an appropriate way; see, e.g., Davis \cite{Davis}, an interview with four industry's top risk executives in September 2006: \emph{``[A] big challenge for us is how to mix the internal data with external data; this is something that is still a big problem because I don't think anybody has a solution for that at the moment.''} Or: \emph{``What can we do when we don't have enough data $[\ldots]$ How do I use a small amount of data when I can have external data with scenario generation? $[\ldots]$ I think it is one of the big challenges for operational risk managers at the moment''}.\\
A ``Toy'' model, based on hierarchical credibility theory, was proposed by Bühl\-mann et al.~\cite{BuShWu06} for low frequency high impact operational risk losses exceeding some high threshold. However, this model can be too sensitive to expert opinions used to estimate scaling factors for distribution parameters. In the present framework we introduce a model that is more robust towards expert opinions.\\
%The model estimates frequency and severity distributions of the low frequency large losses in a risk cell using data %across all cells in a bank, expert opinion on differences between risk cells and industry data. However, some model %assumptions, such as independence between risk cells, might be difficult to justify in practice. Also, extension of %the considered Poisson-Pareto case to other distribution types can be problematic.\\
We use Bayesian inference as the statistical technique to incorporate expert opinions into data analysis. There is a broad literature covering Bayesian inference and its applications to the insurance industry and other areas. The method allows for structural modeling of different sources of information. Shevchenko and Wüthrich \cite{ShWu07} described the use of the Bayesian inference approach, in the context of operational risk, for estimation of frequency/severity distributions in a risk cell, where expert opinion \emph{or} external data are used to estimate prior distributions. This allows the combining of \emph{two} data sources: either expert opinion and internal data or external data and internal data.\\
The novelty in this paper is that we develop a Bayesian inference model that allows for combining three sources (internal data, external data and expert opinions) simultaneously. To the best of our knowledge, we have not seen any similar model that copes comprehensively with this task. Moreover, one should note that our framework enlarges the classical Bayesian inference models belonging to the exponential dispersion family with its associated conjugates; see, e.g., Bühlmann and Gisler \cite{BG}, Chapter 2.\\
In Section \ref{Bayes} we develop a suitable method to combine the three types of knowledge in the context of operational risk. In Sections 3 and 4, this framework is used to quantify loss frequency and severity, respectively. Several examples illustrate the quality and the robustness of this quantitative approach for operational risk. In Section 5 we briefly discuss open challenges when aggregating risk cells and estimating risk capital.

\section{Bayesian Inference}
\label{Bayes}
In order to estimate the risk capital of a bank and to fulfill the Basel II requirements, risk managers have to take into account information beyond the (often rare) internal data. This includes relevant external data (industry data) and expert opinions. The aim of this section is to provide some well-founded background to combining these three sources of information. Hereafter we consider one risk cell only.\\
In any risk cell, we model the loss frequency and the loss severity by a distribution (e.g., Poisson for the frequency or Pareto, lognormal etc.~for the severity). For the considered bank, the unknown parameters $\bm{\gamma}_0$ (e.g., the Poisson parameter or the Pareto tail index) of these distributions have to be quantified.\\
A priori, before we have any company specific information, only industry data are available. Hence, the best prediction of our bank specific parameter $\bm{\gamma}_0$ is given by the belief in the available external knowledge such as the provided industry data. This unknown parameter of interest is modeled by a prior distribution (also called structural distribution or risk profile) corresponding to a random vector $\bm{\gamma}$. The parameters of the prior distribution (so-called hyper-parameters) are estimated using data from the whole industry by, e.g., maximum likelihood estimation, as described in Shevchenko and Wüthrich \cite{ShWu07}. If no industry data are available, the prior distribution could come from a ``super expert'' that has an overview over all banks.\\
In our terminology, we treat the true company specific parameter $\bm{\gamma}_0$ as a realization of $\bm{\gamma}$. The random vector $\bm{\gamma}$ plays the role of the underlying parameter set of the whole banking industry sector, whereas $\bm{\gamma}_0$ stands for the unknown underlying parameter set of the bank being considered. Note that $\bm{\gamma}$ is random with known distribution, whereas $\bm{\gamma}_0$ is deterministic but unknown. Due to the variability amongst banks, it is natural to model $\bm{\gamma}$ by a probability distribution.\\
As time passes, internal observations $\bm{X}=(X_1,\ldots,X_K)$ as well as expert opinions $\bm{\vartheta}=(\vartheta^{(1)},\ldots,\vartheta^{(M)})$ about the underlying parameter $\bm{\gamma}_0$ become available. This affects our belief in the distribution of $\bm{\gamma}$ coming from external data only and adjust the prediction of $\bm{\gamma}_0$. The more information on $\bm{X}$ and $\bm{\vartheta}$ we have, the better we are able to predict $\bm{\gamma}_0$. That is, we replace the prior density $\pi(\bm{\gamma})$ by a conditional density of $\bm{\gamma}$ given $\bm{X}$ and $\bm{\vartheta}$.\\
The natural question that arises at this point is: How does this company specific information $\bm{X}$ and $\bm{\vartheta}$ change our view of the underlying parameter $\bm{\gamma}$, i.e., what is the distribution of $\bm{\gamma} | \bm{X}, \bm{\vartheta}$?\\
The Bayesian inference approach yields the canonical theory answering questions of the above type. In order to determine $\bm{\gamma} | \bm{X}, \bm{\vartheta}$ we have to introduce some notation. The joint conditional density of observations and expert opinions given the parameter vector $\bm{\gamma}$ is denoted by
\be
\label{eq:independence}
h(\bm{X},\bm{\vartheta}|\bm{\gamma}) = h_1(\bm{X}|\bm{\gamma}) h_2(\bm{\vartheta}|\bm{\gamma}),
\ee
where $h_1$ and $h_2$ are the conditional densities (given $\bm{\gamma}$) of $\bm{X}$ and $\bm{\vartheta}$, respectively. Thus $\bm{X}$ and $\bm{\vartheta}$ are assumed to be conditionally independent given $\bm{\gamma}$.
\begin{rmks}
\begin{itemize}
\item Notice that, in this way, we naturally combine external data $\bm{\gamma}$ with internal data $\bm{X}$ and expert opinion $\bm{\vartheta}$.
\item In classical Bayesian inference (as it is used, e.g., in actuarial science), one usually combines only two sources of information. The novelty in this paper is that we combine three sources simultaneously using an appropriate structure, i.e., equation~(\ref{eq:independence}).
\item (\ref{eq:independence}) is quite a reasonable assumption: Assume that the true bank specific parameter is $\bm{\gamma}_0$. Then (\ref{eq:independence}) says that the experts in this bank estimate $\bm{\gamma}_0$ (by their opinion $\bm{\vartheta}$) independently of the internal observations. This makes sense if the experts specify their opinions regardless of the data observed.
\end{itemize}
\end{rmks}
We further assume that observations as well as expert opinions are conditionally independent and identically distributed (i.i.d.), given $\bm{\gamma}$, so that
\begin{eqnarray}
h_1(\bm{X}|\bm{\gamma}) &=& \prod_{k=1}^{K} f_1(X_k | \bm{\gamma}),\\
h_2(\bm{\vartheta}|\bm{\gamma}) &=& \prod_{m=1}^{M} f_2(\vartheta^{(m)} | \bm{\gamma}),
\end{eqnarray}
where $f_1$ and $f_2$ are the marginal densities of a single observation and a single expert opinion, respectively.
We have assumed that all expert opinions are identically distributed, but this can be generalized easily to expert opinions having different distributions.\\
The unconditional parameter density $\pi(\bm{\gamma})$ is called the \emph{prior} density, whereas the conditional parameter density $\widehat{\pi}(\bm{\gamma}| \bm{X},\bm{\vartheta})$ is called the \emph{posterior} density. Let $h(\bm{X},\bm{\vartheta})$ denote the unconditional joint density of observations $\bm{X}$ and expert opinions $\bm{\vartheta}$. Then it follows from Bayes' Theorem that
\be
h(\bm{X},\bm{\vartheta}|\bm{\gamma}) \pi(\bm{\gamma}) = \widehat{\pi}(\bm{\gamma}| \bm{X},\bm{\vartheta}) h(\bm{X},\bm{\vartheta}).
\ee
Note that the unconditional density $h(\bm{X},\bm{\vartheta})$ does not depend on $\bm{\gamma}$ and, thus, the posterior density is given by
\be
\label{aposterioridistribution}
\widehat{\pi}(\bm{\gamma}| \bm{X},\bm{\vartheta}) \propto \pi(\bm{\gamma})
\prod_{k=1}^{K} f_1(X_k | \bm{\gamma})
\prod_{m=1}^{M} f_2(\vartheta^{(m)} | \bm{\gamma}),
\ee
where ``$\propto$'' stands for ``is proportional to'' with the constant of proportionality independent of the parameter vector $\bm{\gamma}$. For the purposes of operational risk it is used to estimate the full predictive distribution of future losses.\\
Equation (\ref{aposterioridistribution}) can be used in a general set-up, but it is convenient to find some \emph{conjugate} prior distributions such that the prior and the posterior distribution have a similar type, or where, at least, the posterior distribution can be calculated analytically.
%Our aim is to find distributions with densities $\pi,f_1,f_2$, so that we can explicitly calculate the a posteriori density $\widehat{\pi}$ given in (\ref{aposterioridistribution}).
\begin{dfn}[Conjugate Prior Distribution]
Let $F$ denote the class of density functions $h(\bm{X},\bm{\vartheta}|\bm{\gamma})$, indexed by $\bm{\gamma}$. A class $U$ of prior densities $\pi(\bm{\gamma})$ is said to be a \emph{conjugate family} for $F$ if the posterior density
$\widehat{\pi}(\bm{\gamma}| \bm{X},\bm{\vartheta}) \propto \pi(\bm{\gamma}) h(\bm{X},\bm{\vartheta}|\bm{\gamma})$ also belongs to the class $U$ for all $h \in F$ and $\pi \in U$.
\end{dfn}
Conjugate distributions are very useful in practice and will be used consistently throughout this paper. At this point, we also refer to Bühlmann and Gisler \cite{BG}, Section 2.5. In general, the posterior distribution cannot be calculated analytically but can be estimated numerically for instance by the Markov Chain Monte Carlo method; see, e.g., Peters and Sisson \cite{PeSi} or Gilks et al.~\cite{Gilks}.

\section{Loss Frequency}
\label{section:claimnumberprocess}
\subsection{Combining internal data and expert opinions with external data}
\begin{model}[Poisson-Gamma-Gamma]
\label{model:PoissonGammaGamma}
Assume that bank $i$ has a scaling factor $V_i$, $1 \leq i \leq I$, for the frequency in a specified risk cell (e.g., it can be a product of economic indicators such as the gross income, the number of transactions, the number of staff, etc.). We choose the following model for the loss frequency for operational risk of a risk cell in bank $i$:
\begin{itemize}
\item[a)] Let $\Lambda_i \sim \Gamma (\alpha_0, \beta_0)$ be a Gamma distributed random variable with shape parameter $\alpha_0>0$ and scale parameter $\beta_0>0$, which are estimated from (external) market data. That is, the density of $\Gamma (\alpha_0, \beta_0)$, $x^{\alpha_0-1} e^{-x/\beta_0} / (\beta_0^{\alpha_0} \Gamma(\alpha_0))$ $(x>0)$, plays the role of $\pi(\bm{\gamma})$ in (\ref{aposterioridistribution}).
\item[b)] The number of losses of bank $i$ in year $k$, $1 \leq k \leq K_i$, are assumed to be conditionally i.i.d., given $\Lambda_i$, Poisson distributed with frequency $V_i \Lambda_i$, i.e., $N_{i,1}, \ldots, N_{i,K_i}|\Lambda_i \stackrel{\textnormal{i.i.d.}}{\sim} \textnormal{Pois}(V_i \Lambda_i)$. That is, $f_1(\cdot|\Lambda_i)$ in (\ref{aposterioridistribution}) corresponds to the density of a $\textnormal{Pois}(V_i \Lambda_i)$ distribution.
\item[c)] We assume that bank $i$ has $M_i$ experts with opinions $\vartheta_i^{(m)}$, $1 \leq m \leq M_i$, about the company specific intensity parameter $\Lambda_i$ with $\vartheta_i^{(m)} | \Lambda_i \stackrel{\textnormal{i.i.d.}}{\sim} \Gamma (\xi_i,\frac{\Lambda_i}{\xi_i})$, where $\xi_i$ is a known parameter. That is, $f_2(\cdot|\Lambda_i)$ corresponds to the density of a $\Gamma (\xi_i,\frac{\Lambda_i}{\xi_i})$ distribution.
\end{itemize}
\end{model}
\begin{rmks}
\begin{itemize}
\item In the sequel, we only look at a single bank $i$ and therefore we could drop the index $i$. However, we refrain from doing so in order to highlight the fact that we do not consider the whole banking industry, but only a single bank.
\item The parameters $\alpha_0$ and $\beta_0$ in Model Assumptions \ref{model:PoissonGammaGamma} a) are called hyper-parameters (parameters for parameters); see, e.g., Bühlmann and Gisler \cite{BG}, p.~38. These parameters are estimated using the maximum likelihood method or the method of moments; see for instance Shevchenko and Wüthrich \cite{ShWu07}, Section 5 and Appendix~B.
\item In Model Assumptions \ref{model:PoissonGammaGamma} c) we assume
\be
\mathds{E}[\vartheta_i^{(m)}|\Lambda_i]=\Lambda_i, \quad 1 \leq m \leq M_i,
\ee
that is, expert opinions are unbiased. A possible bias might only be recognized by the regulator, as he alone has the overview of the whole market.
\end{itemize}
\end{rmks}
Note that the \emph{coefficient of variation} of the conditional expert opinion $\vartheta_i^{(m)}|\Lambda_i$ of company $i$ is $\textnormal{Vco}(\vartheta_i^{(m)}|\Lambda_i)= (\textnormal{var}(\vartheta_i^{(m)}|\Lambda_i))^{1/2} / \mathds{E}[\vartheta_i^{(m)}|\Lambda_i] = 1/\sqrt{\xi_i}$, and thus is independent of $\Lambda_i$. This means that $\xi_i$, which characterizes the uncertainty in the expert opinions, is independent of the true bank specific $\Lambda_i$. For simplicity, we have assumed that all experts have the same conditional coefficient of variation and thus have the same credibility. Moreover, this allows for the estimation of $\xi_i$ within each company $i$, e.g., by $\widehat{\xi}_i= (\widehat{\mu}_i / \widehat{\sigma}_i)^2$ with
\be
\label{eq:xiestimate}
\widehat{\mu}_i = \frac{1}{M_i} \sum_{m=1}^{M_i} \vartheta_i^{(m)} \quad \textnormal{and} \quad
\widehat{\sigma}_i^2 = \frac{1}{M_i - 1} \sum_{m=1}^{M_i} (\vartheta_i^{(m)} - \widehat{\mu}_i)^2, \quad M_i \geq 2.
\ee
%We should distinguish between the random variable $\vartheta_i^{(m)}$ and the estimation $\widehat{\vartheta}_i^{(m)} \in \mathds{R}$. This is neglected here in order to not overload the notation.\\
In a more general framework the parameter $\xi_i$ can be estimated, e.g., by maximum likelihood. If the credibility differs among the experts, then $\vartheta_i^{(m)}$ and $\textnormal{Vco}(\vartheta_i^{(m)}|\Lambda_i)$ should be estimated for all $m$, $1 \leq m \leq M_i$. This may often be a (too) challenging issue in practice.
\begin{rmks}
\begin{itemize}
\item $\Lambda_i$ is the risk characteristic of a risk cell in bank $i$. A priori, before we have any observations, the banks are all the same, i.e., $\Lambda_i$ is i.i.d. Observations and expert opinions modify this characteristic $\Lambda_i$ according to the actual experience in company $i$, which gives different posteriors $\Lambda_i|N_{i,1},\ldots,N_{i,K_i},\vartheta_i^{(1)},\ldots,\vartheta_i^{(M_i)}$.
\item This model can be extended to a model where one allows for more flexibility in the expert opinions. For convenience, we prefer that experts are conditionally i.i.d., given $\Lambda_i$. This has the advantage that there is only one parameter, $\xi_i$, that needs to be estimated.
\end{itemize}
\end{rmks}
Using the notation from Section \ref{Bayes}, we calculate the posterior density of $\Lambda_i$, given the losses up to year $K_i$ and the expert opinion of $M_i$ experts. We introduce the following notation for the loss database and the expert knowledge of bank~$i$:
\begin{eqnarray*}
\bm{N}_i &=& (N_{i,1},\ldots,N_{i,K_i}),\\
\bm{\vartheta}_i &=& (\vartheta_i^{(1)},\ldots,\vartheta_i^{(M_i)}).
\end{eqnarray*}
Here and in what follows, we denote arithmetic means by
\be
\overline{N}_i = \frac{1}{K_i} \sum_{k=1}^{K_i} N_{i,k}, \quad \overline{\vartheta}_i = \frac{1}{M_i} \sum_{m=1}^{M_i} \vartheta_i^{(m)}, \quad \textnormal{etc.}
\ee
The posterior density $\widehat{\pi}$ is given by the following theorem.
\begin{thm}
\label{GIGThm}
Under Model Assumptions \ref{model:PoissonGammaGamma}, the posterior density of $\Lambda_i$, given loss information $\bm{N_i}$ and expert opinion $\bm{\vartheta_i}$, is given by
\be
\label{GIG}
\widehat{\pi}_{\Lambda_i}(\lambda_i | \bm{N_i}, \bm{\vartheta_i}) = \frac{(\omega/\phi)^{(\nu+1)/2}}{2 K_{\nu+1}(2 \sqrt{\omega \phi})} \lambda_i^\nu e^{-\lambda_i \omega - \lambda_i^{-1} \phi},
\ee
with
\begin{eqnarray}
\label{eq:nuomegaphi}
\nu &=& \alpha_0 - 1 - M_i \xi_i + K_i \overline{N_i}, \nonumber\\
\omega &=& V_i K_i + \frac{1}{\beta_0},\\
\phi &=& \xi_i M_i \overline{\vartheta_i},\nonumber
\end{eqnarray}
and
\be
\label{def:modifiedBessel3}
K_{\nu +1}(z) = \frac{1}{2} \int_0^{\infty} u^\nu e^{-z(u+1/u)/2}\textnormal{d}u.
\ee
\end{thm}
$K_{\nu}(z)$ is called a modified Bessel function of the third kind; see for instance Abramowitz and Stegun \cite{AbSt}, p.~375.
\begin{Proof}%[Proof of Theorem \ref{GIGThm}]
Set $\alpha_i = \xi_i$ and $\beta_i = \lambda_i/\xi_i$. Model Assumptions \ref{model:PoissonGammaGamma} applied to (\ref{aposterioridistribution}) yield
\begin{eqnarray}
\widehat{\pi}_{\Lambda_i}(\lambda_i | \bm{N_i}, \bm{\vartheta_i})
&\propto& \lambda_i^{\alpha_0-1} e^{-\lambda_i/\beta_0}
\prod_{k=1}^{K_i} e^{-V_i \lambda_i} \frac{(V_i \lambda_i)^{N_{i,k}}}{N_{i,k}!}
\prod_{m=1}^{M_i} \frac{(\vartheta_i^{(m)}/\beta_i)^{\alpha_i-1}}{\beta_i} e^{-\vartheta_i^{(m)}/\beta_i} \nonumber\\
&\propto& \lambda_i^{\alpha_0-1} e^{-\lambda_i/\beta_0}
\prod_{k=1}^{K_i} e^{-V_i \lambda_i} \lambda_i^{N_{i,k}}
\prod_{m=1}^{M_i} (\xi_i/\lambda_i)^{\xi_i}  e^{-\vartheta_i^{(m)} \xi_i /\lambda_i} \nonumber\\
&\propto& \lambda_i^{\alpha_0-1-M_i \xi_i + K_i \overline{N}_i} \exp{\left(-\lambda_i \left(V_i K_i + \frac{1}{\beta_0} \right) -\frac{1}{\lambda_i} \xi_i M_i \overline{\vartheta}_i \right)}.
\label{formula:numberposteriori}
\end{eqnarray}
\end{Proof}
\begin{rmks}
\begin{itemize}
\item A distribution with density (\ref{GIG}) is referred to as the generalized inverse Gaussian distribution GIG$(\omega,\phi,\nu)$. This is a well-known distribution with many applications in finance and risk management; see McNeil et al.~\cite{MFE}. The GIG has been analyzed by many authors. A discussion is found, e.g., in J\o{}rgensen \cite{Joergensen}. The GIG belongs to the popular class of subexponential distributions; see Embrechts \cite{Em83} for a proof and Embrechts et al.~\cite{EKM} for a detailed treatment of subexponential distributions. The GIG with $\nu \leq 1$ is a first hitting time distribution for certain time-homogeneous processes; see for instance J\o{}rgensen \cite{Joergensen}, Chapter 6. In particular, the (standard) inverse Gaussian (i.e., the GIG with $\nu=-3/2$) is known by financial practitioners as the distribution function determined by the first passage time of a Brownian motion. Algorithms for generating realizations from a GIG are provided by Atkinson \cite{Atkinson82} and Dagpunar \cite{Dagpunar89}; see also McNeil et al.~\cite{MFE} and Appendix \ref{GIGalgorithm} below.
\item Unlike in the classical Poisson-Gamma case of combining two sources of information (see Shevchenko and Wüthrich \cite{ShWu07}, Bühlmann and Gisler \cite{BG}), we obtain in (\ref{formula:numberposteriori}) a more complicated posterior distribution $\widehat{\pi}$, which involves in the exponent both $\lambda_i$ and $1/\lambda_i$. Note that expert opinions enter via the term $1/\lambda_i$ only. We give some basic properties of the GIG distribution below.
\item Observe that the classical exponential dispersion family (EDF) with associated conjugates (see Bühlmann and Gisler \cite{BG}, Chapter 2.5) allows for a natural extension to GIG-like distributions. In this sense the GIG distributions enlarge the classical Bayesian inference theory on the exponential dispersion family.
\end{itemize}
\end{rmks}
\noindent
For our purposes it is interesting to observe how the posterior density transforms when new data from a newly observed year arrive. Let $\nu_k$, $\omega_k$ and $\phi_k$ denote the parameters for the observations $(N_{i,1},\ldots,N_{i,k})$ after $k$ accounting years. Implementation of the update processes is then given by the following equalities (assuming that expert opinions do not change).
\begin{update}
Year $k$ $\rightarrow$ year $k+1$:
\begin{eqnarray}
\nu_{k+1} &=& \nu_k + N_{i,k+1},\nonumber\\
\omega_{k+1} &=& \omega_k + V_i,\\
\phi_{k+1} &=& \phi_k.\nonumber
\end{eqnarray}
\end{update}
Obviously, the information update process has a very simple form and only the parameter $\nu$ is affected by the new observation $N_{i,k+1}$. The posterior density (\ref{formula:numberposteriori}) does not change its type every time new data arrive and hence, is easily calculated.\\
The moments of a GIG cannot be given in a closed form by elementary functions. However, for $\alpha \geq 1$, all moments are given in terms of Bessel functions:
\be
\label{moments}
\mathds{E}[\Lambda_i^\alpha | \bm{N_i}, \bm{\vartheta_i}] = \left(\frac{\phi}{\omega}\right)^{\alpha/2} \frac{K_{\nu + 1 + \alpha}(2 \sqrt{\omega \phi})}{K_{\nu + 1}(2 \sqrt{\omega \phi})}.
\ee
A useful notation is the following:
\be
\label{Rfunction}
R_{\nu}(z) = \frac{K_{\nu + 1}(z)}{K_{\nu}(z)}.
\ee
Then it follows for the posterior expected number of losses
\be
\label{expectedvalue}
\mathds{E}[\Lambda_i| \bm{N_i}, \bm{\vartheta_i}] = \sqrt{\frac{\phi}{\omega}}R_{\nu + 1}(2 \sqrt{\omega \phi}),
\ee
and for the higher moments
\be
\mathds{E}[\Lambda_i^\alpha| \bm{N_i}, \bm{\vartheta_i}] = \left( {\frac{\phi}{\omega}} \right)^{\alpha/2} \prod_{k=1}^\alpha R_{\nu + k}(2 \sqrt{\omega \phi}), \quad \alpha=2,3,\ldots
\ee
We are clearly interested in robust prediction of the bank specific Poisson parameter and thus the Bayesian estimator (\ref{expectedvalue}) is a promising candidate within this operational risk framework. The examples below show that, in practice, (\ref{expectedvalue}) outperforms other classical estimators. To interpret (\ref{expectedvalue}) in more detail, we make use of asymptotic properties. Here and throughout the paper, $f(x) \sim g(x)$, $x \rightarrow a$, means that $\displaystyle{\lim_{x \rightarrow a} \frac{f(x)}{g(x)}=1}$. Lemma \ref{AsymptoticsOfR} in Appendix \ref{Proofs} basically says that $R_{\nu^2}(2 \nu) \sim \nu$ is asymptotically linear for $\nu \rightarrow \infty$. This is the key in the proof of Theorem \ref{AsymptoticsOfExpectation} and yields a full asymptotic interpretation of the Bayesian estimator (\ref{expectedvalue}).
\begin{thm}
\label{AsymptoticsOfExpectation}
Under Model Assumptions \ref{model:PoissonGammaGamma}, the following asymptotic relations hold $\mathds{P}$-almost surely:
\begin{enumerate}
\item[a)] Assume, given $\Lambda_i=\lambda_i,$ $N_{i,k} \stackrel{\textnormal{i.i.d.}}{\sim}$ Pois$(V_i \lambda_i)$ and $\vartheta_i^{(m)} \stackrel{\textnormal{i.i.d.}}{\sim} \Gamma(\xi_i,\lambda_i/\xi_i)$.\\
For $K_i \rightarrow \infty:$ $\mathds{E}[\Lambda_i| \bm{N_i}, \bm{\vartheta_i}] \rightarrow \mathds{E}[N_{i,k}|\Lambda_i=\lambda_i]/V_i = \lambda_i$. %\mathds{E}[N_{i,k}|\Lambda_i]/V_i$ $\mathds{P}$-a.s.
\item[b)] For $\textnormal{Vco}(\vartheta_i^{(m)} | \Lambda_i) \rightarrow 0:$ $\mathds{E}[\Lambda_i| \bm{N_i}, \bm{\vartheta_i}] \rightarrow \vartheta_i^{(m)}$, $m=1,\ldots,M_i$.
%(or $M_i \rightarrow \infty$)
\item[c)] Assume, given $\Lambda_i=\lambda_i,$ $N_{i,k} \stackrel{\textnormal{i.i.d.}}{\sim}$ Pois$(V_i \lambda_i)$ and $\vartheta_i^{(m)} \stackrel{\textnormal{i.i.d.}}{\sim} \Gamma(\xi_i,\lambda_i/\xi_i)$.\\
For $M_i \rightarrow \infty:$ $\mathds{E}[\Lambda_i| \bm{N_i}, \bm{\vartheta_i}] \rightarrow \mathds{E}[\vartheta_i^{(m)}|\Lambda_i=\lambda_i] = \lambda_i$.
\item[d)] For $\textnormal{Vco}(\vartheta_i^{(m)} | \Lambda_i) \rightarrow \infty,$ $m=1,\ldots,M_i:$\\ $\mathds{E}[\Lambda_i| \bm{N_i}, \bm{\vartheta_i}] \rightarrow \frac{1}{V_i K_i \beta_0 + 1}\mathds{E}[\Lambda_i] + \left( 1 - \frac{1}{V_i K_i \beta_0 + 1} \right) \overline{N}_i/V_i.$ %\mathds{E}[N_{i,k}|\Lambda_i]/V_i$ $\mathds{P}$-a.s.
\item[e)] For $\mathds{E}[\Lambda_i]=$ constant and $\textnormal{Vco}(\Lambda_i) \rightarrow 0:$ $\mathds{E}[\Lambda_i| \bm{N_i}, \bm{\vartheta_i}] \rightarrow \mathds{E}[\Lambda_i]$.
\end{enumerate}
\end{thm}
\begin{Proof}
See Appendix \ref{ProofOfAsymptoticsOfExpectation}.
\end{Proof}
Theorem \ref{AsymptoticsOfExpectation} yields a natural interpretation of the posterior density (\ref{GIG}) and its expected value (\ref{expectedvalue}). As the number of observations increases, we give more weight to them and in the limit $K_i \rightarrow \infty$ (case a) we completely believe in the observations $N_{i,k}$ and we neglect a priori information and expert opinion. On the other hand, the more the coefficient of variation of the expert opinions decreases, the more weight is given to them (case b). In Model \ref{model:PoissonGammaGamma}, we assume experts to be conditionally independent. In practice, however, even for $\textnormal{Vco}(\vartheta_i^{(m)}|\Lambda_i) \rightarrow 0$, the variance of $\overline{\vartheta_i}|\Lambda_i$ cannot be made arbitrarily small when increasing the number of experts, as there is always a positive covariance term due to positive dependence between experts. Since we predict random variables, we never have ``perfect diversification'', that is, in practical applications we would probably question property c.\\
Conversely, if experts become less credible in terms of having an increasing coefficient of variation, our model behaves as if the experts do not exist (case d). The Bayes estimator is then a weighted sum of prior and posterior information with appropriate credibility weights. This is the classical credibility result obtained from Bayesian inference on the exponential dispersion family with two sources of information; see Shevchenko and Wüthrich~\cite{ShWu07}, Formula (12).\\
Of course, if the coefficient of variation of the prior distribution (i.e., of the whole banking industry) vanishes, the external data are not affected by internal data and expert opinion (case e).\\
In this sense, Theorem \ref{AsymptoticsOfExpectation} shows that our model behaves exactly as we would expect and require in practice. Thus, we have good reasons to believe that it provides an adequate model to combine internal observations with relevant external data and expert opinions, as required by many risk managers.\\
Note that one can even go further and generalize the results from this section in a natural way to a Poisson-Gamma-GIG model, i.e., where the prior distribution is a GIG. Then the posterior distribution is again a GIG (see also Model Assumptions \ref{model:heavytailedGIG} below).

\subsection{Implementation and practical application}
In this section we apply the above theory to a concrete example. The Bayesian estimator (\ref{expectedvalue}) derived above is easily  implemented in practice. The following example extends the example displayed in Figure 1 in Shevchenko and Wüthrich \cite{ShWu07}.
\begin{bsp}
\label{example:frequeny}
Assume that external data (e.g., provided by external databases or regulator) estimate the parameter of the loss frequency (i.e., the Poisson parameter $\Lambda$) which has a Gamma distribution $\Lambda \sim \Gamma (\alpha_0,\beta_0)$ as $\mathds{E}[\Lambda]=\alpha_0 \beta_0 = 0.5$ and $\mathds{P}[0.25 \leq \Lambda \leq 0.75]= 2/3$. Then, the parameters of the prior Gamma distribution are $\alpha_0 \approx 3.407$ and $\beta_0 \approx 0.147$; see Shevchenko and Wüthrich \cite{ShWu07}, Section 4.1.\\
Now, we consider one particular bank $i$:
\begin{enumerate}
\item[i)] One expert says that $\vartheta$ is estimated by $\widehat{\vartheta}=0.7$. For simplicity, we consider in this example one single expert only and hence, the coefficient of variation is not estimated using (\ref{eq:xiestimate}), but given a priori, e.g., by the regulator: $\textnormal{Vco}(\vartheta|\Lambda) = (\textnormal{var}(\vartheta|\Lambda))^{1/2} / \mathds{E}[\vartheta|\Lambda] = 0.5$, i.e., $\xi=4$.
\item[ii)] The observations of the annual number of losses are given as follows (sampled from a Poisson distribution with parameter $\lambda = 0.6$; this is the dataset used in Shevchenko and Wüthrich \cite{ShWu07}):

\begin{center}
%\vspace{0.5cm}
\begin{tabular}{lccccccccccccccc}
\hline
Year $i$& 1 &2& 3& 4& 5& 6& 7& 8& 9& 10&11&12&13&14&15 \\
$N_i$& 0&0&0&0&1&0&1&1&1&0&2&1&1&2&0 \\
%\hline
%Claim $k$&11& 12& 13& 14& 15\\
%Severity $X_k$&1.59 & 1.35 & 1.91 & 1.23 & 1.03\\
\hline
\end{tabular}
\end{center}
\end{enumerate}
\noindent
This means that a priori we have a frequency parameter distributed as $\Gamma(\alpha_0,\beta_0)$ with mean $\alpha_0 \beta_0 = 0.5$. The true parameter for this institution is $\lambda = 0.6$, i.e., it does worse than the average institution. However, our expert has an even worse opinion of his institution, namely $\widehat{\vartheta} = 0.7$.

%\vspace{0cm}
\begin{figure}[!ht]
  \begin{center}
  \setcaptionwidth{12cm}
    \includegraphics[width=0.95\textwidth]{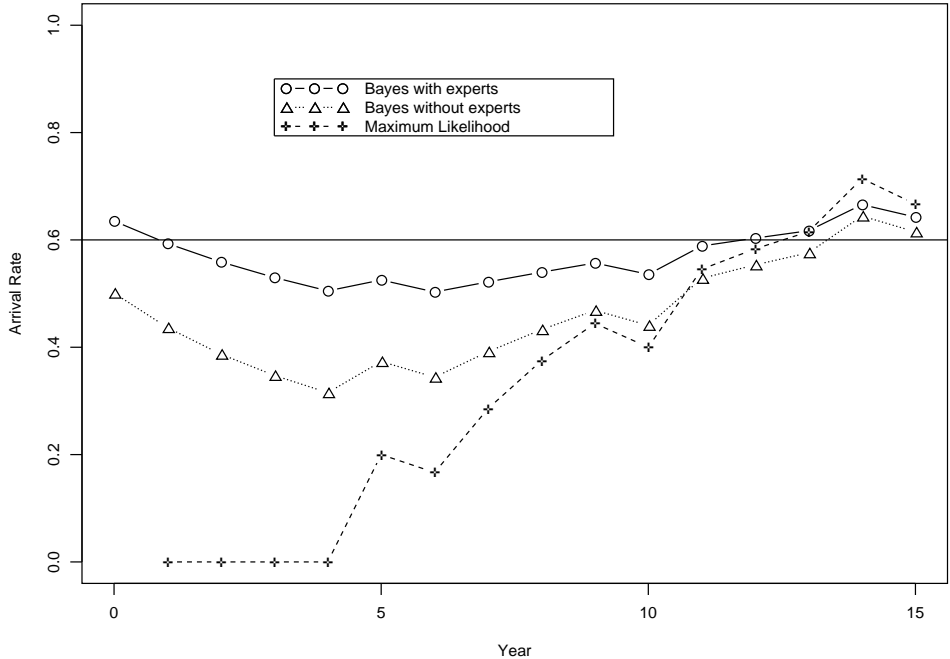}
    \caption{The Bayes estimator $(\lambda_k)_{k=0,\ldots,15}$ includes internal data simulated from Poisson(0.6), external data $\Lambda$ with $\mathds{E}[\Lambda]=0.5$ and expert opinion $\widehat{\vartheta}=0.7$ $(\circ)$. It is compared with the Bayes estimator $\lambda_k^{\textnormal{SW}}$ proposed in Shevchenko and Wüthrich \cite{ShWu07} $(\triangle)$ and the classical maximum likelihood estimator $(+)$.}
    \label{fig:bayesclaimnumber1}
  \end{center}
\end{figure}
\noindent
We compare the pure maximum likelihood estimator $\lambda_k^{\textnormal{MLE}} = \frac{1}{k} \sum_{i=1}^{k} N_i$ and the Bayesian estimator
\be
\lambda_k^{\textnormal{SW}} = \mathds{E}[\Lambda|N_1,\ldots,N_k],
\ee
proposed in Shevchenko and Wüthrich \cite{ShWu07} (without expert opinion) with the Bayesian estimator derived in formula (\ref{expectedvalue}), including expert opinion:
\be
\lambda_k = \mathds{E}[\Lambda|N_1,\ldots,N_k,\vartheta].
\ee
The results are plotted in Figure \ref{fig:bayesclaimnumber1}. The estimator (\ref{expectedvalue}) shows a much more stable behavior around the true value $\lambda=0.6$, due to the use of the prior information (market data) and the expert opinions. Given adequate expert opinions, the Bayesian estimator (\ref{expectedvalue}) clearly outperforms the other estimators, particularly if only a few data points are available.

\begin{figure}[!ht]
  \begin{center}
  \setcaptionwidth{12cm}
   \includegraphics[width=0.95\textwidth]{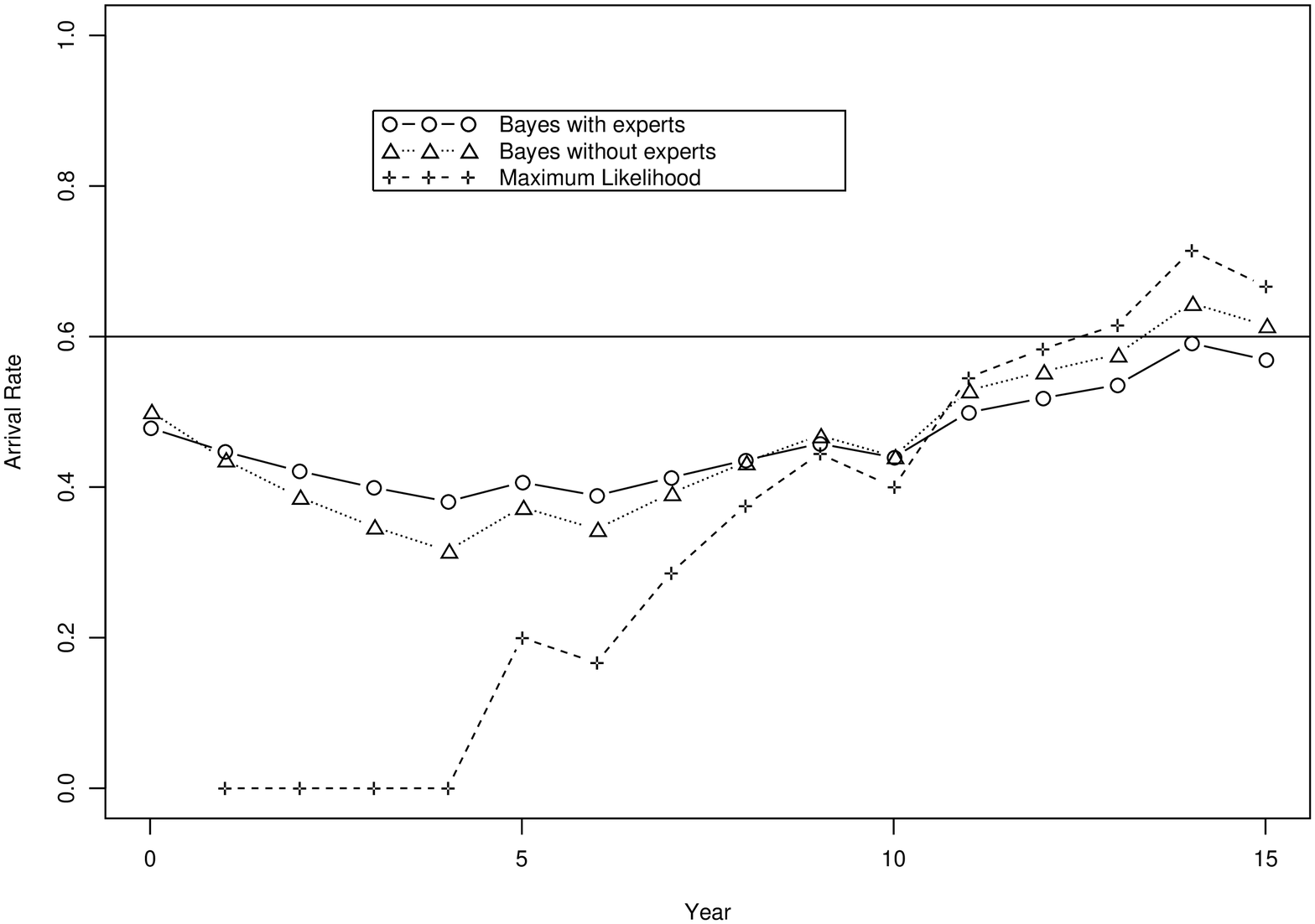}
   \caption{The same estimators as in Figure \ref{fig:bayesclaimnumber1} are displayed, but where the expert underestimates the true $\lambda=0.6$ by $\widehat{\vartheta}=0.4$.}
    \label{fig:bayesclaimnumber2}
  \end{center}
\end{figure}
\noindent
One could think that this is only the case when the experts' estimates are appropriate. However, even if experts fairly under- (or over-) estimate the true parameter $\lambda$, the method presented in this paper performs better for our dataset than the other mentioned methods, when a few data are available. In Figure \ref{fig:bayesclaimnumber2} we display the same estimators, but where the experts' opinion is $\widehat{\vartheta}=0.4$, which clearly underestimates the true expected value 0.6.\\
In Figure \ref{fig:bayesclaimnumber1} $\lambda_k$ gives better estimates when compared to $\lambda_k^{\textnormal{SW}}$. Observe that also in Figure \ref{fig:bayesclaimnumber2} $\lambda_k$ gives more appropriate estimates than $\lambda_k^{\textnormal{SW}}$. Though the expert is too optimistic, $\lambda_k$ manages to correct $\lambda_k^{\textnormal{MLE}}$ $(k \leq 10)$, which is clearly too low.
\end{bsp}
This example yields a typical picture observed in numerical experiments that demonstrates that the Bayes estimator (\ref{expectedvalue}) is often more suitable and stable than maximum likelihood estimators based on internal data only.
\begin{rmk}
Note that in this example the prior distribution as well as the expert opinion do not change over time. However, as soon as new information is available or when new risk management tools are in place, the corresponding parameters may be easily adjusted.
\end{rmk}

\subsection{Alternative estimator using the mode}
Instead of calculating the mean of the GIG$(\nu,\omega,\phi)$ as we did in the estimator (\ref{expectedvalue}), we could use the \emph{mode} of the distribution, i.e., the point where the density function is maximum. The mode of a GIG differs only slightly from the expected value for large $|\nu|$. In particular, one proves, e.g., that for $X \sim \textnormal{GIG}(\nu,\omega,\phi)$ we have
\be
\textnormal{mode}(X) \sim \mathds{E}[X] \qquad \textnormal{for} \quad \nu \rightarrow \infty.
\ee
The mode of a GIG$(\nu,\omega,\phi)$ is easily calculated by
\be
\frac{\partial}{\partial x} x^\nu e^{-(\omega x + \phi / x)} = 0.
\ee
Hence,
\be
\textnormal{mode}(X) = \frac{1}{2 \omega} (\nu + \sqrt{\nu^2 + 4 \omega \phi}),
\ee
which gives us a good approximation to the mean for large $\nu$. Thus, we have
\be
\textnormal{mode}(\Lambda_i | \bm{N_i}, \bm{\vartheta_i}) = \frac{\nu}{2 \omega}
\left( 1 + \textnormal{sign}(\nu) \sqrt{1+\frac{4 \omega \phi}{\nu^2}} \right),
\ee
where $\nu$, $\omega$, and $\phi$ are given by equations (\ref{eq:nuomegaphi}). Due to $\frac{4 \omega \phi}{\nu^2} \rightarrow 0$ for $K_i \rightarrow \infty$, $M_i \rightarrow \infty$, $M_i \rightarrow 0$ or $\xi \rightarrow 0$, we approximate $\sqrt{1+2x} \approx 1+x$, $x \rightarrow 0$, and hence
\be
\label{eq:mode}
\textnormal{mode}(\Lambda_i | \bm{N_i}, \bm{\vartheta_i}) \approx \frac{\nu}{2 \omega} \mathds{1}_{\{ \nu \geq 0 \}} +\frac{\phi}{|\nu|}.
\ee
With (\ref{eq:mode}) we again get the results from Theorem \ref{AsymptoticsOfExpectation} in an elementary manner avoiding Bessel functions.

\section{Loss Severities}
In the previous section we presented a method to quantify the operational risk loss frequency. We now turn to quantification of the severity distribution for operational risk. This is done in this section for different types of subexponential models.

\subsection{Lognormal model (Model 1 for severities)}
\label{subsection:lognormal}
\begin{model}[Lognormal-normal-normal]
\label{model:subexp}
Let us assume the following severity model for operational risk of a risk cell in bank $i$, $1 \leq i \leq I$:
\begin{itemize}
\item[a)] Let $\Delta_i \sim \mathcal{N} (\mu_0, \sigma_0)$ be a normally distributed random variable with parameters $\mu_0, \sigma_0$, which are estimated from (external) market data, i.e., $\pi(\bm{\gamma})$ in (\ref{aposterioridistribution}) is the density of $\mathcal{N}(\mu_0,\sigma_0)$.
\item[b)] The losses $k=1,\ldots,K_i$ from institution $i$ are assumed to be conditionally (on $\Delta_i$) i.i.d.~lognormally distributed: $X_{i,1}, \ldots, X_{i,K_i}|\Delta_i \stackrel{\textnormal{i.i.d.}}{\sim} \textnormal{LN}(\Delta_i,\sigma_i)$, where $\sigma_i$ is assumed known. That is, $f_1(\cdot|\Delta_i)$ in (\ref{aposterioridistribution}) corresponds to the density of a $\textnormal{LN}(\Delta_i,\sigma_i)$ distribution.
\item[c)] We assume that bank $i$ has $M_i$ experts with opinions $\vartheta_i^{(m)}$, $1 \leq m \leq M_i$, about the parameter $\Delta_i$ with $\vartheta_i^{(m)} | \Delta_i \stackrel{\textnormal{i.i.d.}}{\sim} \mathcal{N} (\mu_i=\Delta_i,\widetilde{\sigma}_i=\xi_i)$, where $\xi_i$ is a parameter estimated using expert opinion data. That is, $f_2(\cdot|\Delta_i)$ corresponds to the density of a $\mathcal{N}(\Delta_i,\xi_i)$ distribution.
\end{itemize}
\end{model}
\begin{rmks}
\begin{itemize}
\item For $M_i \geq 2$, the parameter $\xi_i$ is, e.g., estimated by the standard deviation of $\vartheta_i^{(m)}$:
\be
\xi_i = \left( \frac{1}{M_i - 1} \sum_{m=1}^{M_i} (\vartheta_i^{(m)} - \overline{\vartheta_i})^2 \right)^{1/2}.
\ee
\item The hyper-parameters $\mu_0$ and $\sigma_0$ are estimated from market data, e.g., by maximum likelihood estimation or by the method of moments.
\item In practice one often uses an ad hoc estimate for $\sigma_i$, $1 \leq i \leq I$, which usually is based on expert opinion only. However one could think of a Bayesian approach for $\sigma_i$, but then an analytical formula for the posterior distribution in general does not exist. The posterior distribution needs then to be calculated for example by the Markov Chain Monte Carlo method; see again Peters and Sisson \cite{PeSi} or Gilks et al.~\cite{Gilks}.
\end{itemize}
\end{rmks}
Under Model Assumption \ref{model:subexp}, the posterior density is given by
\begin{eqnarray}
\widehat{\pi}_{\Delta_i}(\delta_i | \bm{X_i}, \bm{\vartheta_i})
&\propto& \frac{1}{\sigma_0 \sqrt{2 \pi}} \exp \left( - \frac{(\delta_i-\mu_0)^2}{2 \sigma_0^2} \right)
\prod_{k=1}^{K_i}  \frac{1}{\sigma_i \sqrt{2 \pi}} \exp \left( - \frac{(\log X_{i,k}-\delta_i)^2}{2 \sigma_i^2} \right)\nonumber\\
&& \prod_{m=1}^{M_i}  \frac{1}{\widetilde{\sigma}_i \sqrt{2 \pi}} \exp \left( - \frac{(\vartheta_i^{(m)}-\delta_i)^2}{2 \widetilde{\sigma}_i^2} \right)\nonumber\\
&\propto& \exp \left[-\left(\frac{(\delta_i-\mu_0)^2}{2 \sigma_0^2} + \sum_{k=1}^{K_i} \frac{1}{2\sigma_i^2} (\log X_{i,k}-\delta_i)^2 \right. \right. \nonumber\\
&& \left. \left. + \sum_{m=1}^{M_i} \frac{1}{2\xi_i^2} (\vartheta_i^{(m)}-\delta_i)^2 \right) \right] \nonumber \\
%&\propto&
%\exp \left[ -\frac{1}{2\widehat{\sigma}} \left( \mu_i -( (1-\omega_1-\omega_2) \mu_0 + \omega_1 \frac{1}{K_i} \sum_{k=1}^{K_i} \log X_k  \right. \right.\nonumber\\
%&& \left. \left. + \omega_2\frac{1}{M_i} \sum_{m=1}^{M_i} \vartheta_i^{(m)} ) \right)^2 \right]\nonumber,
&\propto& \exp \left[ -\frac{(\delta_i - \widehat{\mu})^2}{2 \widehat{\sigma}^2} \right],
\end{eqnarray}
with
\be
\widehat{\sigma}^2 = \left( \frac{1}{\sigma_0^2} + \frac{K_i}{\sigma_i^2} + \frac{M_i}{\xi_i^2} \right)^{-1},
\ee
and
\be
\widehat{\mu} = \widehat{\sigma}^2 \cdot \left( \frac{\mu_0}{\sigma_0^2} + \frac{1}{\sigma_i^2} \sum_{k=1}^{K_i} \log X_{i,k} + \frac{1}{\xi_i^2} \sum_{m=1}^{M_i} \vartheta_i^{(m)} \right).
\ee
In summary we have the following theorem.
\begin{thm}
Under Model Assumptions \ref{model:subexp} and with the notation $\overline{\log X_i} = \frac{1}{K_i} \sum_{k=1}^{K_i} \log X_{i,k}$, the posterior distribution of $\Delta_i$, given loss information $\bm{X_i}$ and expert opinion $\bm{\vartheta_i}$, is a normal distribution $\mathcal{N}(\widehat{\mu},\widehat{\sigma})$ with
\be
\widehat{\sigma}^2 = \left( \frac{1}{\sigma_0^2} + \frac{K_i}{\sigma_i^2} + \frac{M_i}{\xi_i^2} \right)^{-1},
\ee
and
\be
\label{eq:subexpmean}
\widehat{\mu} = \mathds{E}[\Delta_i| \bm{X_i}, \bm{\vartheta_i}] = \omega_1 \mu_0 + \omega_2 \overline{\log X_i} +
\omega_3 \overline{\vartheta_i}.
\ee
The credibility weights are $\omega_1 = \widehat{\sigma}^2 / \sigma_0^2$, $\omega_2 = \widehat{\sigma}^2 K_i / \sigma_i^2$ and $\omega_3 = \widehat{\sigma}^2 M_i / \xi_i^2$.
\end{thm}
This theorem yields a natural interpretation of the considered model. The estimator $\widehat{\mu}$ in (\ref{eq:subexpmean}) weights the internal and external data as well as the expert opinion in an appropriate manner. Observe that under Model Assumptions \ref{model:subexp} we can explicitly calculate the mean of the posterior distribution. This is different from the frequency model in Section \ref{section:claimnumberprocess}. That is, we have an exact calculation and for the interpretation of the terms we do not rely on an asymptotic theorem as in Theorem \ref{AsymptoticsOfExpectation}. However, interpretation of the terms is exactly the same as in Theorem \ref{AsymptoticsOfExpectation}. The more credible the information, the higher is the credibility weight in (\ref{eq:subexpmean}). Hence, again, this theorem shows that our model is appropriate for combining internal observations, relevant external data and expert opinions.

\subsection{Pareto model (Model 2 for severities)}
\label{section:pareto}
\begin{model}[Pareto-Gamma-Gamma]
\label{model:heavytailed}
Let us assume the following severity model for a particular operational risk cell of bank $i$, $1 \leq i \leq I$:
\begin{itemize}
\item[a)] Let $\Gamma_i \sim \Gamma (\alpha_0, \beta_0)$ be a Gamma distributed random variable with parameters $\alpha_0, \beta_0$, which are estimated from (external) market data, i.e., $\pi(\bm{\gamma})$ in (\ref{aposterioridistribution}) is the density of a $\Gamma(\alpha_0,\beta_0)$ distribution.
\item[b)] The losses $k=1,\ldots,K_i$ from institution $i$ are assumed to be conditionally (on $\Gamma_i$) i.i.d.~Pareto distributed: $X_{i,1}, \ldots, X_{i,K_i}|\Gamma_i \stackrel{\textnormal{i.i.d.}}{\sim} \textnormal{Pareto}(\Gamma_i,L_i)$, where the threshold $L_i\geq0$ is assumed to be known and fixed. That is, $f_1(\cdot|\Gamma_i)$ in (\ref{aposterioridistribution}) corresponds to the density of a Pareto$(\Gamma_i,L_i)$ distribution.
\item[c)] We assume that bank $i$ has $M_i$ experts with opinions $\vartheta_i^{(m)}$, $1 \leq m \leq M_i$, about the parameter $\Gamma_i$ with $\vartheta_i^{(m)} | \Gamma_i \stackrel{\textnormal{i.i.d.}}{\sim} \Gamma(\alpha_i=\xi_i,\beta_i=\Gamma_i/\xi_i)$, where $\xi_i$ is a parameter estimated using expert opinion data; see (\ref{eq:xiestimate}). That is, $f_2(\cdot|\Gamma_i)$ corresponds to the density of a $\Gamma(\xi_i,\Gamma_i/\xi_i)$ distribution.
\end{itemize}
\end{model}
Under Model Assumptions \ref{model:heavytailed}, the posterior density is given by
\begin{eqnarray}
\widehat{\pi}_{\Gamma_i}(\gamma_i | \bm{X_i}, \bm{\vartheta_i})
&\propto& \gamma_i^{\alpha_0-1} e^{-\gamma_i/\beta_0}
\prod_{k=1}^{K_i} \frac{\gamma_i}{L_i} \left(\frac{X_{i,k}}{L_i}\right)^{-(\gamma_i+1)}
\prod_{m=1}^{M_i} \frac{(\vartheta_i^{(m)}/\beta_i)^{\alpha_i-1}}{\beta_i}  e^{-\vartheta_i^{(m)}/\beta_i}\nonumber\\
&\propto& \gamma_i^{\alpha_0-1-M_i \xi_i +K_i} \exp \left[ -\gamma_i \left( \frac{1}{\beta_0} + \sum_{k=1}^{K_i} \log \frac{X_{i,k}}{L_i}  \right) -\frac{1}{\gamma_i} \xi_i M_i \overline{\vartheta_i} \right].
\label{eq:heavyGIGdensity}
\end{eqnarray}
Hence, again, the posterior distribution is a GIG and it has the nice property that the term $\gamma_i$ in the exponent in (\ref{eq:heavyGIGdensity}) is only affected by the internal observations, whereas the term $1/\gamma_i$ is driven by the expert opinions.
\begin{thm}
\label{theorem:heavytailedAposteriori}
Under Model Assumptions \ref{model:heavytailed}, the posterior density of $\Gamma_i$, given loss information
$\bm{X_i}$ and expert opinion $\bm{\vartheta_i}$, is given by
\be
\label{formula:aposterioriParetoGammaGamma}
\widehat{\pi}_{\Gamma_i}(\gamma_i | \bm{X_i}, \bm{\vartheta_i}) = \frac{(\omega/\phi)^{(\nu+1)/2}}{2 K_{\nu+1}(2 \sqrt{\omega \phi})} \gamma_i^\nu e^{-\gamma_i \omega - \gamma_i^{-1} \phi},
\ee
with
\begin{eqnarray}
\nu &=& \alpha_0-1-M_i \xi_i +K_i,\nonumber\\
\label{eq:nuomegaphiGamma}
\omega &=& \frac{1}{\beta_0} + \sum_{k=1}^{K_i} \log \frac{X_{i,k}}{L_i},\\
\phi &=& \xi_i M_i \overline{\vartheta_i}\nonumber.
\end{eqnarray}
\end{thm}
It seems natural to generalize this result by substituting the prior Gamma distribution by a GIG as follows.

\begin{model}[Pareto-Gamma-GIG]
\label{model:heavytailedGIG}
Let us assume the following severity model for a particular operational risk cell of bank $i$, $1 \leq i \leq I$:
\begin{itemize}
\item[a)] Let $\Gamma_i \sim \textnormal{GIG}(\nu_0, \omega_0, \phi_0)$ be a generalized inverse Gaussian distributed random variable with parameters $\nu_0,$ $\omega_0,$ $\phi_0$, which are estimated from (external) market data, i.e., $\pi(\bm{\gamma})$ in (\ref{aposterioridistribution}) is the density of a GIG$(\nu_0,\omega_0,\phi_0)$ distribution.
\item[b)] The losses $k=1,\ldots,K_i$ from bank $i$ are assumed to be conditionally (on $\Gamma_i$) i.i.d. Pareto distributed: $X_{i,1}, \ldots, X_{i,K_i}|\Gamma_i \stackrel{\textnormal{i.i.d.}}{\sim} \textnormal{Pareto}(\Gamma_i,L_i)$, where the threshold $L_i\geq0$ is assumed to be known and fixed. That is, $f_1(\cdot|\Gamma_i)$ in (\ref{aposterioridistribution}) corresponds to the density of a Pareto$(\Gamma_i,L_i)$ distribution.
\item[c)] We assume that bank $i$ has $M_i$ experts with opinions $\vartheta_i^{(m)}$, $1 \leq m \leq M_i$, about the parameter $\Gamma_i$ with $\vartheta_i^{(m)} | \Gamma_i \stackrel{\textnormal{i.i.d.}}{\sim} \Gamma(\alpha_i=\xi_i,\beta_i=\Gamma_i/\xi_i)$, where $\xi_i$ is a parameter estimated using expert opinion data. That is, $f_2(\cdot|\Gamma_i)$ corresponds to the density of a $\Gamma(\xi_i,\Gamma_i/\xi_i)$ distribution.
\end{itemize}
\end{model}
Under Model Assumptions \ref{model:heavytailedGIG}, the a posteriori density $\widehat{\pi}_{\Gamma_i}(\gamma_i | \bm{X_i}, \bm{\vartheta_i})$ is given by (\ref{formula:aposterioriParetoGammaGamma})
%\begin{eqnarray*}
%\widehat{\pi}_{\Gamma_i}(\gamma_i | \bm{X_i}, \bm{\vartheta_i})
%&\propto& \gamma_i^{\gamma_0} e^{-\gamma_i \omega_0 - \gamma_i^{-1} \phi_0}
%\cdot \prod_{k=1}^{K_i} \frac{\gamma_i}{L} \left(\frac{X_k}{L}\right)^{-(\gamma_i+1)}
%\cdot \prod_{m=1}^{M_i} \frac{(\vartheta_i^{(m)}/\beta_i)^{\alpha_i-1}}{\beta_i} e^{-\vartheta_i^{(m)}/\beta_i}\\
%&\propto& \gamma_i^{\gamma_0-M_i \xi_i +K_i} \exp \left[ -\gamma_i \left( \omega_0 + \sum_{k=1}^{K_i} \log %\frac{X_k}{L}  \right) -\frac{1}{\gamma_i} \left( \phi_0 + \xi_i \sum_{m=1}^{M_i} \vartheta_i^{(m)} \right) \right].
%\end{eqnarray*}
with
\begin{eqnarray}
\nu &=& \nu_0-M_i \xi_i +K_i,\nonumber\\
\label{eq:nuomegaphiGIG}
\omega &=& \omega_0 + \sum_{k=1}^{K_i} \log \frac{X_{i,k}}{L_i},\\
\phi   &=& \phi_0  + \xi_i M_i \overline{\vartheta_i}\nonumber.
\end{eqnarray}
Hence, again, the posterior distribution is given by a GIG.
%\begin{thm}
%Under Model Assumptions \ref{model:heavytailedGIG}, the a posteriori density of $\Gamma_i$ given loss information
%$\bm{X_i}$ and expert opinion $\bm{\vartheta_i}$ is given by
%\be
%\widehat{\pi}_{\Gamma_i}(\gamma_i | \bm{X_i}, \bm{\vartheta_i}) = \frac{(\omega/\phi)^{(\gamma+1)/2}}{2 K_{\gamma+1}(2 %\sqrt{\omega \phi})} \gamma_i^\gamma e^{-\gamma_i \omega - \lambda_i^{-1} \phi},
%\ee
%with
%\begin{eqnarray*}
%\gamma &=& \gamma_0+K_i,\\
%\omega &=& \omega_0 + \sum_{k=1}^{K_i} \log \frac{X_k}{L},\\
%\phi &=& \phi_0 + \xi_i \sum_{m=1}^{M_i} \vartheta_i^{(m)}.
%\end{eqnarray*}
%\end{thm}
Note that for $\phi_0=0$, the GIG is a Gamma distribution and hence we are in the Pareto-Gamma-Gamma situation of Model \ref{model:heavytailed}.\\
The following theorem gives us a natural interpretation of the Bayesian estimator
\be
\label{BayesEstimatorHeavy}
\mathds{E}[\Gamma_i| \bm{X_i}, \bm{\vartheta_i}] = \sqrt{\frac{\phi}{\omega}}R_{\nu + 1}(2 \sqrt{\omega \phi}).
\ee
Denote the maximum likelihood estimator of the Pareto tail index $\Gamma_i$ by
\be
\gamma_i^{\textnormal{MLE}}=\frac{K_i}{\sum_{k=1}^{K_i} \log \frac{X_{i,k}}{L_i}}.
\ee
Then, completely analogous to Theorem \ref{AsymptoticsOfExpectation} we obtain the following theorem.
\begin{thm}
\label{thm:asympt}
Under Model Assumptions \ref{model:heavytailed} and \ref{model:heavytailedGIG}, the following asymptotic relations hold $\mathds{P}$-almost surely:
\begin{enumerate}
\item[a)] Assume, given $\Gamma_i=\gamma_i,$ $X_{i,k} \stackrel{\textnormal{i.i.d.}}{\sim}$ \textnormal{Pareto}$(\gamma_i , L_i)$ and $\vartheta_i^{(m)} \stackrel{\textnormal{i.i.d.}}{\sim} \Gamma(\xi_i,\gamma_i/\xi_i)$.\\
For $K_i \rightarrow \infty:$ $\mathds{E}[\Gamma_i| \bm{X_i}, \bm{\vartheta_i}] \rightarrow \mathds{E}[X_{i,k}|\Gamma_i=\gamma_i]/V_i = \gamma_i.$
\item[b)] For $\textnormal{Vco}(\vartheta_i^{(m)} | \Gamma_i) \rightarrow 0:$ $\mathds{E}[\Gamma_i| \bm{X_i}, \bm{\vartheta_i}] \rightarrow \vartheta_i^{(m)},$ $m=1,\ldots,M_i$.
\item[c)] Assume, given $\Gamma_i=\gamma_i,$ $X_{i,k} \stackrel{\textnormal{i.i.d.}}{\sim}$ \textnormal{Pareto}$(\gamma_i , L_i)$ and $\vartheta_i^{(m)} \stackrel{\textnormal{i.i.d.}}{\sim} \Gamma(\xi_i,\gamma_i/\xi_i)$.\\
For $M_i \rightarrow \infty:$ $\mathds{E}[\Gamma_i| \bm{X_i}, \bm{\vartheta_i}] \rightarrow \mathds{E}[\vartheta_i^{(m)}|\Gamma_i=\gamma_i] = \gamma_i.$
\item[d)] For $\textnormal{Vco}(\vartheta_i^{(m)} | \Gamma_i) \rightarrow \infty,$ $m=1,\ldots,M_i:$\\
$\mathds{E}[\Gamma_i| \bm{X_i}, \bm{\vartheta_i}] \rightarrow \left( 1 - \frac{K_i \beta_0}{\gamma_i^{\textnormal{MLE}} + K_i \beta_0} \right) \mathds{E}[\Gamma_i]
+ \frac{K_i \beta_0}{\gamma_i^{\textnormal{MLE}} + K_i \beta_0} \gamma_i^{\textnormal{MLE}}$.
\item[e)] For $\mathds{E}[\Gamma_i]=$ constant and $\textnormal{Vco}(\Gamma_i) \rightarrow 0:$ $\mathds{E}[\Gamma_i| \bm{X_i}, \bm{\vartheta_i}] \rightarrow \mathds{E}[\Gamma_i]$.
\end{enumerate}
\end{thm}
\begin{rmks}
\begin{itemize}
\item Theorem \ref{thm:asympt} basically says that the higher the precision of a particular source of risk information, the higher its corresponding credibility weight. This means that we obtain the same interpretations as for Theorem \ref{AsymptoticsOfExpectation} and Formula (\ref{eq:subexpmean}).
\item Observe that in Section \ref{section:claimnumberprocess} and Section \ref{subsection:lognormal} we have applied Bayesian inference to the expected values of the Poisson and the normal distribution, respectively. However, Bayesian inference is much more general, and basically, can be applied to any reasonable parameter. In this Section \ref{section:pareto} it is, e.g., applied to the Pareto tail index.
\item Observe that Model Assumptions \ref{model:heavytailed} and \ref{model:heavytailedGIG} lead to an infinite mean model because the Pareto parameter $\Gamma_i$ can be less than one with positive probability. For finite mean models, the range of possible $\Gamma_i$ has to be restricted to $\Gamma_i > 1$. This does not impose difficulties; for more details we refer the reader to Shevchenko and Wüthrich \cite{ShWu07}, Section 3.4.
\end{itemize}
\end{rmks}

%\begin{Proof}
%Using Lemma \ref{BesselAsymptotics} and Lemma \ref{AsymptoticsOfR} in Section \ref{Proofs}, we have
%\begin{enumerate}
%\item[a)]
%$\sqrt{\frac{\phi}{\omega}}R_{\gamma+1}(2\sqrt{\omega \phi})
%\sim \sqrt{\frac{\phi}{\omega}} R_{K_i}(2\sqrt{(1/\beta_0 + K_i/\gamma_i^{\textnormal{MLE}}) \phi})
%\sim \sqrt{\frac{K_i \gamma_i^{\textnormal{MLE}}}{\omega}}
%\sim \gamma_i^{\textnormal{MLE}}.$
%\item[b)] See proof of Theorem \ref{AsymptoticsOfExpectation} b).
%\item[c)]
%$\sqrt{\frac{\phi}{\omega}}R_{\gamma+1}(2\sqrt{\omega \phi}) \sim \frac{\Gamma(\gamma+2)}{\omega \Gamma(\gamma+1)}
%= \frac{\gamma+1}{\omega}
%\sim \frac{\alpha_0 + K_i}{1/\beta_0 + K_i/\gamma_i^{\textnormal{MLE}}}$\\
%$=\left( 1 - \frac{K_i \beta_0}{\gamma_i^{\textnormal{MLE}} + K_i \beta_0} \right) \alpha_0 \beta_0
%+ \frac{K_i \beta_0}{\gamma_i^{\textnormal{MLE}} + K_i \beta_0} \gamma_i^{\textnormal{MLE}}$.
%\end{enumerate}
%\end{Proof}

\subsection{Implementation and practical application}
%The Model \ref{model:heavytailed} yields (again) an easy to implement Bayesian estimator (\ref{BayesEstimatorHeavy}) with $\nu,$ $\omega,$ $\phi$ from Theorem \ref{theorem:heavytailedAposteriori}.
Note that the update process of (\ref{eq:nuomegaphiGamma}) and (\ref{eq:nuomegaphiGIG}) has again a simple linear form when new information arrives. The posterior density (\ref{formula:aposterioriParetoGammaGamma}) does not change its type every time a new observation arrives. In particular, only the parameter $\omega$ is affected by a new observation.
\begin{update}
Loss $k$ $\rightarrow$ loss $k+1$:
\begin{eqnarray}
\nu_{k+1} &=& \nu_k + 1,\nonumber\\
\omega_{k+1} &=& \omega_k + \log \frac{X_{i,k+1}}{L_i},\\
\phi_{k+1} &=& \phi_k.\nonumber
\end{eqnarray}
\end{update}
The following example shows the simplicity and robustness of the estimator developed.
\begin{bsp}
Assume that a bank would like to model its risk severity by a Pareto distribution with tail index $\Gamma$. The regulator provides external prior data, saying that $\Gamma \sim \Gamma(\alpha_0,\beta_0)$ with $\alpha_0=4$ and $\beta_0=9/8$, i.e., $\mathds{E}[\Gamma]=4.5$ and $\textnormal{Vco}(\Gamma)=0.5$. The bank has one expert opinion $\widehat{\vartheta}=3.5$ with $\textnormal{Vco}(\vartheta|\Gamma)=0.5$, i.e., $\xi=4$. We then observe the following losses (sampled from a Pareto$(\alpha=4,L=1)$ distribution); see also Figure~\ref{fig:bayesseverity1}:
%\begin{table}[!h]
\begin{center}
%\vspace{0.5cm}
\begin{tabular}{lccccccccccccccc}
\hline
Loss index $i$& 1 &2& 3& 4& 5& 6& 7& 8& 9& 10 \\
Severity $X_i$& 1.17 & 1.29 & 1.00 & 1.55 & 2.66 & 1.02 & 1.28 & 1.10 & 1.06 & 1.02\\
\hline
Loss index $i$&11& 12& 13& 14& 15\\
Severity $X_i$&1.59 & 1.35 & 1.91 & 1.23 & 1.03\\
\hline

%claim & 1.17 & 1.29 & 1.00 & 1.55 & 2.66 & 1.02 & 1.28 & 1.10 & 1.06 & 1.02 & 1.59 & 1.35 & 1.91 & 1.23 & 1.03
%year $k$& 1 &2& 3& 4& 5& 6& 7& 8& 9& 10 \\
%claim $X_k$& 1.173 & 1.285 & 1.003 & 1.552 & 2.661 & 1.024 & 1.275 & 1.101 & 1.062 & 1.015\\
%&11& 12& 13& 14& 15\\
%&1.592 & 1.347 & 1.914 & 1.231 & 1.028

\end{tabular}
\end{center}
%\end{table}

\begin{figure}[!ht]
  \begin{center}
  \setcaptionwidth{12cm}
    \includegraphics[width=0.95\textwidth]{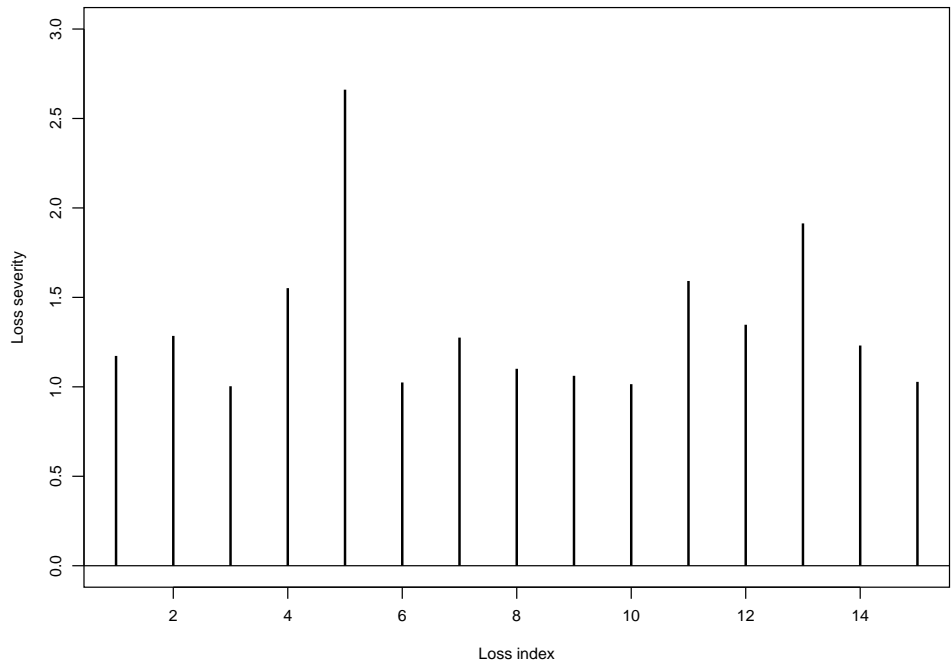}
    \caption{15 loss severities sampled from a Pareto$(\alpha=4,L=1)$ distribution.}
    \label{fig:bayesseverity1}
  \end{center}
\end{figure}

\begin{figure}[!ht]
  \begin{center}
  \setcaptionwidth{12cm}
   \includegraphics[width=0.95\textwidth]{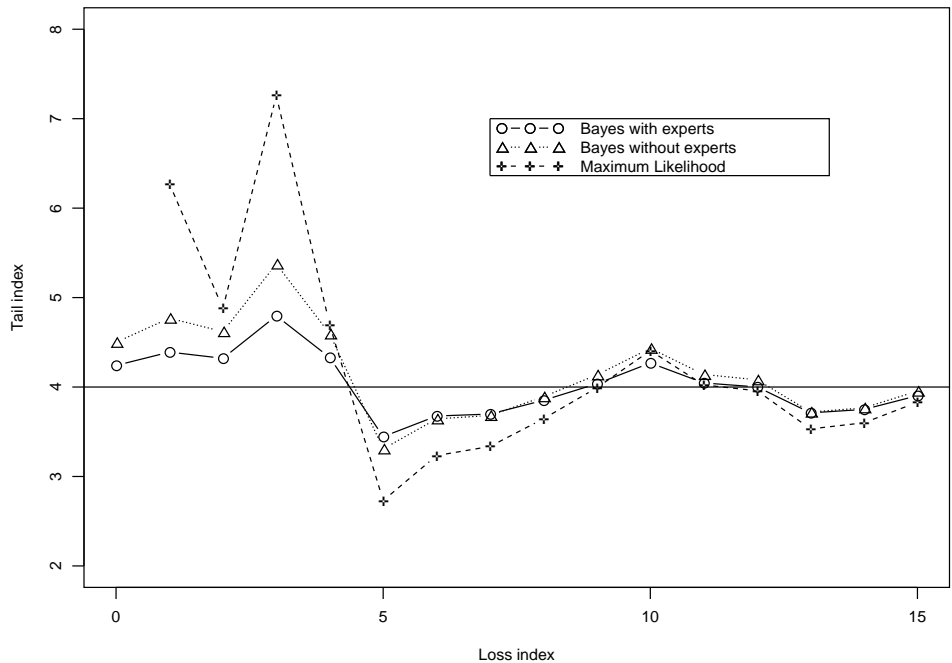}
   \caption{The Bayes estimator including expert opinions ($\circ$) is compared with the Bayes estimator without expert opinions ($\triangle$) and with the maximum likelihood estimator ($+$).}
    \label{fig:bayesseverity2}
  \end{center}
\end{figure}

\noindent
In Figure \ref{fig:bayesseverity2} we compare the Bayes estimator
\be
\gamma_k = \mathds{E}[\Gamma|X_1,\ldots,X_k,\vartheta],
\ee
given by (\ref{BayesEstimatorHeavy}) with the estimator proposed in Shevchenko and Wüthrich \cite{ShWu07} without expert opinions
\be
\gamma_k^{\textnormal{SW}} = \mathds{E}[\Gamma|X_1,\ldots,X_k],
\ee
and the classical maximum likelihood estimator
\be
\gamma_k^{\textnormal{MLE}}= \frac{k}{\sum_{i=1}^{k} \log \frac{X_i}{L}}.
\ee
Figure \ref{fig:bayesseverity2} shows the high volatility of the maximum likelihood estimator, for small numbers $k$. It is very sensitive to newly arriving losses. However, the estimator proposed in this paper shows a much more stable behavior around the true value $\alpha = 4$, most notably when a few data points are available.
\end{bsp}
This example also shows that when modeling severities of operational risk, Bayesian inference is a suitable method to combine different sources of information. The consideration of relevant external data and well-specified expert opinions stabilizes and smoothens the estimator in an appropriate way.

\section{Total loss distribution and risk capital estimates}
In the preceding sections we have described how the parameters of the distributions are estimated. According to the Basel II requirements (see BIS \cite{Basel}) the final bank capital should be calculated as a sum of the risk measures in the risk cells if the bank's model cannot account for correlations between risks accurately. If this is the case, then one needs to calculate VaR for each risk cell separately and sum VaRs over risk cells to estimate the total bank capital. Adding quantiles over the risk cells to find the quantile of the total loss distribution is sometimes too conservative. It is equivalent to the assumption of perfect dependence between risks.\\
The calculation of VaR (taking into account parameter uncertainty) for each risk cell can, in view of the previous sections, easily be done using a simulation approach described in Shevchenko and Wüthrich \cite{ShWu07}, Section~6. Simulation procedures for independent risk cells and in the case of dependence between risks are also described in Shevchenko and Wüthrich \cite{ShWu07} and thus we refrain from commenting further on this issue.\\
However, reasonable aggregation is still an open challenging problem that needs further investigation. The choice of appropriate dependence structures is crucial and determines the amount of diversification. In the general case, when no information about the dependence structure is available, Embrechts and Puccetti \cite{EmPu06} work out bounds for aggregated operational risk capital; for further issues regarding aggregation we would like to refer to Embrechts et al.~\cite{ENW07}.

\section{Conclusion}
In this paper we propose a novel approach that allows for combining three data sources: internal data, external data and expert opinions. The approach is based on the Bayesian inference method. It is applied to the quantification of the frequency and severity distributions in operational risk, where there is a strong need for such a method to meet the Basel II regulatory requirements.\\
The method is based on specifying prior distributions for the parameters of the frequency and severity distributions using industry data. Then, the prior distributions are weighted by the actual observations and expert opinions from the bank to estimate the posterior distributions of the model parameters. These are used to estimate the annual loss distribution for the next reporting year. Estimation of low frequency risks using this method has several appealing features such as: stable estimators, simple calculations (in the case of conjugate priors), and the ability to take expert opinions and industry data into account. This method also allows for calculation of VaR with parameter uncertainty taken into account.\\
For convenience we have assumed that expert opinions are i.i.d.~but all formulas can easily be generalized to the case of expert opinions modeled by different distributions.\\
It would be ideal if the industry risk profiles (prior distributions for frequency and severity parameters in risk cells) are calculated and provided by the regulators to ensure consistency across the banks. Unfortunately this may not be realistic at the moment. Banks might thus estimate the industry risk profiles using industry data available through external databases from vendors and consortia of banks. The data quality, reporting and survival biases in external databases are the issues that should be considered in practice but go beyond the purposes of this paper.\\
The approach described is not too complicated and is well suited for operational risk quantification. It has a simple structure, which is beneficial for practical use and can engage the bank risk managers, statisticians and regulators in productive model development and risk assessment. The model provides a framework that can be developed further by considering other distribution types, dependencies between risks and dependence on time.\\
One of the features of the described method is that the variance of the posterior distribution $\widehat{\pi}(\bm{\gamma}|\cdot)$ will converge to zero for a large number of observations. That is, the true values of the risk parameters will be known exactly. However, there are many factors (for example, political, economical, legal, etc.)~changing in time that will not permit for the precise knowledge of the risk parameters. One can model this by limiting the variance of the posterior distribution by some lower levels (say, e.g., 5\%). This has been done in many solvency approaches for the insurance industry; see, e.g., the Swiss Solvency Test, FOPI \cite{SST}, formulas (25)-(26).\\
Although the main impetus motivation for the present paper is an urgent need from operational risk practitioners, the proposed method is also useful in other areas (such as credit risk, insurance, environmental risk, ecology etc.)~where, mainly due to lack of internal observations, a combination of internal data with external data and expert opinions is required.

%-----------------------------------APPENDIX---------------------------------------

\appendix

\section{Generating realizations from a GIG random variable}
\label{GIGalgorithm}
For practical purposes it is required to generate realizations of a random variable $X \sim \textnormal{GIG}(\omega,\phi,\nu)$
with $\omega, \phi >0$. Observe that we need to construct a special algorithm since we can not invert the distribution function analytically. The following algorithm can be found in Dag\-pu\-nar~\cite{Dagpunar89}; see also McNeil et al.~\cite{MFE}:
\begin{algorithm}[Generalized inverse Gaussian]
\begin{enumerate}
\item $\alpha=\sqrt{\omega/\phi}$; $\beta=2 \sqrt{\omega \phi}$,\\
$m=\frac{1}{\beta} \left( \nu + \sqrt{\nu^2+\beta^2} \right)$,\\
$g(y)=\frac{1}{2}\beta y^3-y^2 (\frac{1}{2}\beta m + \nu + 2) + y (\nu m - \frac{\beta}{2}) + \frac{1}{2}\beta m$.
\item Set $y_0 = m$,\\
While $g(y_0) \leq 0$ do $y_0 = 2 y_0$,\\
$y_+$: root of $g$ in the interval $(m,y_0)$,\\
$y_-$: root of $g$ in the interval $(0,m)$.
\item $a = (y_{+}-m) \left(\frac{y_{+}}{m}\right)^{\nu/2} \exp \left( -\frac{\beta}{4} (y_{+}+\frac{1}{y_{+}}-m-\frac{1}{m}) \right)$,\\
$b = (y_{-}-m) \left(\frac{y_{-}}{m}\right)^{\nu/2} \exp \left( -\frac{\beta}{4} (y_{-}+\frac{1}{y_{-}}-m-\frac{1}{m}) \right)$,\\
$c = -\frac{\beta}{4}\left(m+\frac{1}{m}\right) + \frac{\nu}{2} \log(m)$.
\item Repeat $U,V \sim \textnormal{Unif}(0,1)$, $Y=m+a \frac{U}{V} + b \frac{1-V}{U}$,\\
until $Y>0$ and $-\log U \geq - \frac{\nu}{2} \log Y + \frac{1}{4} \beta (Y + \frac{1}{Y}) + c$,\\
Then $X=\frac{Y}{\alpha}$ is GIG$(\omega,\phi,\nu)$; see Dagpunar \cite{Dagpunar89}.
\end{enumerate}
\noindent
To generate a sequence of $n$ realizations from a GIG random variable, step 4 is repeated $n$ times.
\end{algorithm}

\section{Asymptotic results for modified Bessel functions}
\label{Proofs}

Let $K_{\gamma}(z)$ denote the modified Bessel function of the third kind as defined in (\ref{def:modifiedBessel3}).

%\begin{lem}
%\label{BesselAsymptotics}
%For $\gamma \rightarrow \infty$, the following asymptotic relation holds:
%\be
%K_\gamma(z) \sim  \sqrt{\frac{\pi}{2}} 2^\gamma \gamma^{\gamma-1/2} e^{-\gamma} z^{-\gamma}, \qquad \forall z > 0.
%\ee
%\end{lem}
%\begin{Proof}
%See Ismail \cite{Ismail77}.
%\end{Proof}

\begin{lem}
\label{AsymptoticsOfR}
With notation (\ref{Rfunction}), we have the following asymptotic relation for $\nu \rightarrow \infty$, for all $a,b>0$:
\be
R_{b\nu}(a\sqrt{\nu}) \sim \frac{2 b \sqrt{\nu}}{a}.
\ee
\end{lem}

%\begin{Proof}
%Using (\ref{Rfunction}) in the first and Lemma \ref{BesselAsymptotics} for $x \rightarrow \infty$ in the second step, %we have
%\begin{eqnarray*}
%R_{bx^2}(ax) &=& \frac{K_{bx^2+1}(ax)}{K_{bx^2}(ax)}\\
%&\sim& 2 \frac{(bx^2+1)^{(bx^2+1/2)}}{eax(bx^2)^{(bx^2-1/2)}}\\
%&=& \frac{2}{eax} \left( 1+ \frac{1}{bx^2}\right)^{bx^2} \sqrt{bx^2(bx^2+1)} \\
%&\sim& \frac{2bx}{a}, \qquad x \rightarrow \infty.
%\end{eqnarray*}
%\end{Proof}
\begin{Proof}
From Abramowitz and Stegun \cite{AbSt}, Paragraph 9.7.8 and Olver \cite{Olver}, Chapter 4, we may deduce for large $\nu$ and $z \geq 0$
\be
\label{eq:highorderbessel}
K_{\nu}(\nu z) = \sqrt{\frac{\pi}{2 \nu}} \frac{\exp (-\nu \sqrt{1 + z^2})}{(1 + z^2)^{1/4}}
\left( \frac{z}{1+\sqrt{1+z^2}} \right)^{-\nu} \left( 1 + \varepsilon(\nu,z) \right),
\ee
where the error term $\varepsilon(\nu,z)$ is bounded by
\be
\label{eq:error}
|\varepsilon(\nu,z)| \leq \frac{1}{\nu-\nu_0} \int_0^{1/\sqrt{1+z^2}} |u_1'(s)|ds,
\ee
with $\nu_0 = \frac{1}{6 \sqrt{5}} + \frac{1}{12}$ and $u_1(s) = (3s-5s^3)/24$; see Abramowitz and Stegun \cite{AbSt} for details.\\
In (\ref{eq:highorderbessel}) we replace $\nu$ by $b\nu$ and $z$ by $z_1 = a / \sqrt{\nu}$. The error term $\varepsilon(b\nu,a / \sqrt{\nu})$ in (\ref{eq:highorderbessel}) is then vanishing for $\nu \rightarrow \infty$, because the right-hand side of (\ref{eq:error}) tends to 0. Analogously, we replace $\nu$ by $b\nu + 1$ and $z$ by $z_2 = a\sqrt{\nu} / (b\nu + 1)$ and observe that $\varepsilon(b\nu+1,a\sqrt{\nu} / (b\nu + 1))$ tends to 0. Thus, (\ref{eq:highorderbessel}) gives us asymptotic expressions for $K_{b\nu}(a \sqrt{\nu})$ and $K_{b\nu + 1}(a \sqrt{\nu})$. Straightforward calculations then yield
\be
R_{b\nu}(a\sqrt{\nu}) = \frac{K_{b\nu + 1}(a \sqrt{\nu})}{K_{b\nu}(a \sqrt{\nu})} \sim \frac{2}{z_2} \sim \frac{2 b \sqrt{\nu}}{a}, \qquad \nu \rightarrow \infty.
\ee
This completes the proof.
\end{Proof}

\section{Proof of Theorem \ref{AsymptoticsOfExpectation}}
\label{ProofOfAsymptoticsOfExpectation}
\begin{Proof}
With (\ref{expectedvalue}) the proof of this theorem is straightforward, using Lemma \ref{AsymptoticsOfR} in Appendix \ref{Proofs}. The following statements hold $\mathds{P}$-almost surely.
\begin{enumerate}
\item[a)]
$\sqrt{\frac{\phi}{\omega}}R_{\nu+1}(2\sqrt{\omega \phi})
\sim \sqrt{\frac{\phi}{\omega}} R_{K_i \overline{N_i}}(2\sqrt{V_i K_i \phi}) \sim \sqrt{\frac{K_i}{V_i \omega}} \overline{N_i} \sim \mathds{E}[N_{i,k}|\Lambda_i]/V_i$.%, \quad K_i \rightarrow \infty.$
\item[b,c)]
$\sqrt{\frac{\phi}{\omega}}R_{\nu+1}(2\sqrt{\omega \phi}) \sim \sqrt{\frac{\phi}{\omega}} R_{-M_i\xi_i}(2\sqrt{\omega \xi_i M_i \overline{\vartheta_i}})
\sim \sqrt{\frac{\phi}{\omega}} \frac{1}{R_{M_i\xi_i}(2\sqrt{\omega \xi_i M_i \overline{\vartheta_i}})}$\\
$\sim \sqrt{\frac{\phi \overline{\vartheta_i} }{\xi_i M_i}} = \overline{\vartheta_i} = \vartheta_i^{(m)},$ $m=1,\ldots,M_i$.
\item[d)] If $\xi=0$, we are in the Gamma case $\Gamma(\alpha,\beta)$ with $\alpha = \alpha_0 + K_i \overline{N}_i$ and $\beta = \beta_0/(V_i K_i \beta_0 + 1)$. Hence,\\
$\mathds{E}[\Lambda_i| \bm{N_i}, \bm{\vartheta_i}]=\alpha \beta =
\frac{1}{V_i K_i \beta_0 + 1}\mathds{E}[\Lambda_i] + \left( 1 - \frac{1}{V_i K_i \beta_0 + 1} \right) \overline{N}_i/V_i$.
\item[e)]
$\sqrt{\frac{\phi}{\omega}}R_{\nu+1}(2\sqrt{\omega \phi})
\sim \sqrt{\frac{\phi \mathds{E}[\Lambda_i]}{\alpha_0}}R_{\alpha_0}(2\sqrt{\frac{\alpha_0 \phi}{\mathds{E}[\Lambda_i]}})
\sim \mathds{E}[\Lambda_i]$.
\end{enumerate}
\end{Proof}

%Acknowledgments

\vspace{1cm}
\noindent
\textbf{Acknowledgments}\\
\noindent
The authors would like to thank Isaac Meilijson for several fruitful discussions and Paul Embrechts for his useful comments.

%--------------------------------- Bibliography -----------------------------------------------

%\addcontentsline{toc}{chapter}{Bibliography}
\bibliography{bibliography}

\begin{thebibliography}{10}

\bibitem{AbSt}
Abramowitz, M. and Stegun, I.~A. (1965) {\em Handbook of Mathematical
  Functions\/}.
\newblock Dover Publications, New York.

\bibitem{AlGaLe06}
Alderweireld, T., Garcia, J.  and Léonard, L. (2006) A practical operational
  risk scenario analysis quantification. {\em Risk Magazine\/} {\bf
  19\textnormal{(2)}}, 93--95.

\bibitem{Atkinson82}
Atkinson, A.~C. (1982) The simulation of \textnormal{generalized inverse
  Gaussian} and hyperbolic random variables. {\em SIAM Journal of Scientific
  and Statistical Computation\/} {\bf 3}, 502--515.

\bibitem{BG}
Bühlmann, H. and Gisler, A. (2005) {\em A Course in Credibility Theory and its
  Applications\/}.
\newblock Springer, Berlin.

\bibitem{BuShWu06}
Bühlmann, H., Shevchenko, P.~V.  and Wüthrich, M.~V. (2007) A ``toy'' model for
  operational risk quantification using credibility theory. {\em Journal of
  Operational Risk\/} {\bf 2\textnormal{(1)}}, 3--19.

\bibitem{Basel}
BIS (2005) {\em \textnormal{Basel II: International Convergence of Capital
  Measurement and Capital Standards: a revised framework}\/}. Bank for
  International Settlements (BIS), {\tt www.bis.org}.

\bibitem{Cruz}
Cruz, M.~G. (2002) {\em Modeling, Measuring and Hedging Operational Risk\/}.
\newblock Wiley, Chichester.

\bibitem{Dagpunar89}
Dagpunar, J.~S. (1989) An easily implemented \textnormal{generalised inverse
  Gaussian} generator. {\em Communications in Statistics, Simulation and
  Computation\/} {\bf 18}, 703--710.

\bibitem{Davis}
Davis, E. (2006) \textnormal{Theory vs Reality}. {\em OpRisk and Compliance.\/}
  1 September 2006, \\{\tt
  www.opriskandcompliance.com/public/showPage.html?page=345305}.

\bibitem{DEL}
Degen, M., Embrechts, P.  and Lambrigger, D.~D. (2007) The quantitative
  modeling of operational risk: between g-and-h and \textnormal{EVT}.
  \textnormal{ASTIN Bulletin, to appear.}

\bibitem{DuPe06}
Dutta, K. and Perry, J. (2006) {\em \textnormal{A tale of tails: an empirical
  analysis of loss distribution models for estimating operational risk
  capital}\/}. Federal Reserve Bank of Boston, Working Paper No 06-13.

\bibitem{Em83}
Embrechts, P. (1983) A property of the \textnormal{generalized inverse
  Gaussian} distribution with some applications. {\em Journal of Applied
  Probability\/} {\bf 20}, 537--544.

\bibitem{EKM}
Embrechts, P., Klüppelberg, C.  and Mikosch, T. (1997) {\em Modelling Extremal
  Events for Insurance and Finance\/}.
\newblock Springer, Berlin.

\bibitem{ENW07}
Embrechts, P., Ne\v{s}lehov\'a, J.  and Wüthrich, M.~V. (2007) Additivity
  properties for \textnormal{Value-at-Risk} under archimedean dependence and
  heavy-tailedness. \textnormal{Preprint, ETH Zurich}.

\bibitem{EmPu06}
Embrechts, P. and Puccetti, G. (2006) Aggregating risk capital, with an
  application to operational risk. {\em The Geneva Risk and Insurance Review\/}
  {\bf 31\textnormal{(2)}}, 71--90.

\bibitem{SST}
FOPI (2006) {\em \textnormal{Swiss Solvency Test, Technical Document}\/}.
  Federal Office of Private Insurance, Bern. {\tt
  www.bpv.admin.ch/themen/00506/00552}.

\bibitem{Gilks}
Gilks, W.~R., Richardson, S.  and Spiegelhalter, D.~J. (1996) {\em Markov Chain
  Monte Carlo in practice\/}.
\newblock Chapman \& Hall, London.

\bibitem{Joergensen}
J\o{}rgensen, B. (1982) {\em Statistical Properties of the Generalized Inverse
  Gaussian Distribution\/}.
\newblock Springer, New York.

\bibitem{MFE}
McNeil, A.~J., Frey, R.  and Embrechts, P. (2005) {\em Quantitative Risk
  Management: Concepts, Techniques and Tools\/}.
\newblock Princeton University Press, Princeton.

\bibitem{Mo04}
Moscadelli, M. (2004) {\em \textnormal{The modelling of operational risk:
  experiences with the analysis of the data collected by the Basel
  Committee}\/}. Bank of Italy, Working Paper No 517.

\bibitem{Olver}
Olver, F.~W.~J. (1962) {\em Mathematical Tables, vol.~6, Tables for Bessel
  functions of moderate or large orders\/}.
\newblock National Physical Laboratory. Her Majesty's Stationery Office,
  London.

\bibitem{Panjer}
Panjer, H.~H. (2006) {\em Operational Risks: Modeling Analytics\/}.
\newblock Wiley, New York.

\bibitem{PeSi}
Peters, G.~W. and Sisson, S.~A. (2006) Bayesian inference, \textnormal{Monte
  Carlo} sampling and operational risk. {\em Journal of Operational Risk\/}
  {\bf 1\textnormal{(3)}}, 27--50.

\bibitem{ShWu07}
Shevchenko, P.~V. and Wüthrich, M.~V. (2006) The structural modeling of
  operational risk via \textnormal{Bayesian} inference: combining loss data
  with expert opinions. {\em Journal of Operational Risk\/} {\bf
  1\textnormal{(3)}}, 3--26.

\end{thebibliography}
\bibliographystyle{bibstyle}  % acm, plain, alpha, abbrv, astron, agsm
%\nocite{*}
\end{document}